\documentclass[aps,11pt,nofootinbib,preprintnumbers,showpacs]{revtex4}
\pdfoutput=1

\usepackage{amssymb, amsfonts, amsmath, bm}
\usepackage[pdftex]{graphicx,color}
\usepackage{ulem}
\usepackage{wrapfig}
\usepackage{ulem}

\newcommand{\tatevec}[2]{\left(   \begin{array}{c}
                         #1 \\ #2 \end{array}\right) }

\begin{document}

\title{\Large Construction and application of variations 
on the cylindrical gravitational waves of Weber,Wheeler, and Bonnor
}

\author{Takashi Mishima${}^1$\footnote{tmishima@phys.ge.cst.nihon-u.ac.jp} and Shinya Tomizawa${}^{2}$\footnote{tomizawasny@stf.teu.ac.jp}}
\vspace{2cm}
\affiliation{
${}^1$ Laboratory of Physics, College of Science and Technology, Nihon University,
Narashinodai, Funabashi, Chiba 274-8501, Japan\\
${}^2$ Department of Liberal Arts, Tokyo University of Technology, 5-23-22, Nishikamata, Otaku, Tokyo 144-8535, Japan, 
}

\pacs{04.20.Jb, 04.30.-w}
\begin{abstract}
To clarify certain nonlinear properties of strong gravitational field, 
we investigate cylindrically symmetric gravitational waves that are localized as  
regular wave packets in the space of radial and time coordinates. The waves are constructed by applying a certain kind of harmonic mapping method to the seed solutions with linear polarization, which are generalizations of the solution representing a cylindrical gravitational pulse wave discussed by Weber-Wheeler and Bonnor. 
The solutions obtained here, though their form is rather simple, show occurrence of strong mutual conversion between a linear mode and a cross mode apparently. The single localized wave shows the conversion in the vicinity of the symmetric axis 
where the self-interaction is strengthened, and the collision between multiple waves also causes the conversion. 
These phenomena can be thought to be the emergence of genuine nonlinearity that the Einstein gravity holds. 
Finally we discuss a simple, but interesting application of the solutions to the case of the Einstein-Maxwell system. 
 \end{abstract}

\maketitle

\section{Introduction}
\label{sec:1}

Just before the elapse of a hundred years since Einstein predicted the existence of gravitational waves, 
the prediction was directly confirmed at last \cite{Abbotetal2016-1}. Needless to say, the success of the discovery is owing to 
the heroic experimental efforts \cite{Abbotetal2016-2}. At the same time, the success has also been supported by the sustained great theoretical developments of general relativity, directly or indirectly. This epoch-making discovery must open up a new astronomical window in the near future, and also may bring us the clues to unlock the secret of gravitation and spacetime in the Universe. The new progress brought by the discovery will also urge the study based on the exact solutions of 
the Einstein equations (one of most traditional approaches) to enter into a new phase wherein strong nonlinear effects that have not been discovered so far can be found as new observational phenomena in gravitational waves.

\medskip
Actually, however, as pointed out by \cite{Alekseev2016}, most of the traditional studies done so far  
mainly have focused on the geometric analysis of spacetime structure, and comparatively the studies treating dynamical effects of gravitational interaction are not so many. For example, as the main stream of study of general relativity that once existed, we easily come up with the studies made intensively from the 1970s to the 1980s  by using the exact solutions that represent colliding plane waves (for more details, see \cite{book colliding waves} and references therein). Most of the studies were devoted to rather restricted aspects of nonlinearity originating from the Einstein gravity, for example, the global structure of spacetimes or formation of singularities. Relatively recently, several researchers have studied the physical wave phenomena related to nonlinear interaction of gravitational waves itself: the gravitational  Faraday effect~\cite{Piran,tomimatsu}, the time shift phenomena \cite{Dagotto:1991, Griffiths:1997}, and so on (for more details, see \cite{bicak:2000} and references therein). From the present standpoint after the direct discovery of gravitational waves, it is not too much to say that many physically interesting solutions, regardless of whether or not they are known, still remain without being thoroughly investigated in different ways than before.

\medskip
To shed some new light on nonlinear features of gravitational waves, 
in previous works~\cite{Tomizawa:2013soa,Tomizawa:2015zva,Igata:2015oea,Igata:2016}, 
we studied cylindrically symmetric gravitational solitonic waves 
with nonaligned polarizations that are constructed on flat and Levi-Civit\`{a} spacetime  
backgrounds by the inverse scattering method \cite{Belinsky:1979mh, Belinski:2001ph}. 
Through those studies, some interesting behaviors of the waves (time shift, gravitational Faraday effect, etc.) 
were examined. 
Related to the present work, it is noteworthy that the solutions dealt with especially in the second paper \cite{Tomizawa:2015zva} have 
important characteristics to observe the nonlinear features of the waves clearly.  
In summary, the characteristics of these waves are as follows: 
the background where the gravitational interactions occur is a regular flat spacetime;  
the waves are regular and localized on a one-dimensional space of a radial coordinate in a cylindrical coordinate 
system, and are asymptotically well behaved; 
even single waves must become inevitably strong when concentrated on the axis of symmetry at the reflection. 
These characteristics may be advantageous to the deduction of more detailed information about the effects of 
strong gravitational interactions, in comparison with other types of exact solutions like plane symmetric waves 
on flat backgrounds or cosmological backgrounds. 
Actually, the solutions concerned here can be considered to represent regular and localized waves propagating in the one-dimensional space, so that we can take a simple viewpoint of one-dimensional wave phenomena, for example, reflection of a wave at a boundary ({\it i.e.,} a symmetric axis), or collision of an ingoing wave and an outgoing wave. 
Then we just need to analyze the behavior of the one-dimensional wave by tracing the time sequence of 
the so-called cylindrical energy (C-energy) and its derivatives ({\it i.e.,} density and fluxes) \cite{Thorn}. 
To be able to use the well-defined gravitational energies like the C-energy is another advantage of cylindrical symmetric cases. 
The work~\cite{Ashtekar:1997} presented another type of energy that is mathematically established. 
In the rest of this article, however, we use the C-energy only, because both of the two energies show the same behavior qualitatively as shown in the last section.  
We can therefore perform unambiguously the analysis to extract physical characteristics based on 
these quantities.

\medskip
Standing on the above-mentioned viewpoint, we will advance the investigation of nonlinear features of 
gravitational waves.
The solitonic waves considered previously have rich structure, but are rather complicated for systematic 
handling. 
To develop the study, we need a different type of solution. 
Hence we adopt, as research objects in this paper, some variations of 
the  famous exact solution that was discussed by Weber-Wheeler and Bonnor (WWB) 
in the early period 
\cite{weber:1957,bonnor:1957}. 
The original WWB solution represents the cylindrical gravitational wave that can be considered to be localized 
on the one-dimensional space.
The WWB solution belongs to a class of Einstein-Rosen-type solutions\cite{Einstein:1937}, which is given by solving a linear wave 
equation directly, and further has a simple expression described with elementary functions. 
So the WWB solution has been used to clarify physical features of gravitational waves for a long time 
(for example, see \cite{book Exact Space-Times}). Recently, the cylindrical wave solution was used to study dragging effects by gravitational waves~\cite{Bicak:2008gn,LyndenBell:2008gr,Bicak:2012zz}.
However,  in principle, the Einstein-Rosen-type solutions have only a linear polarization mode ({\it i.e.,} $+$ mode), so that genuine nonlinearity that originates from the interaction of independent two modes ({\it i.e.,} $+$ and $\times$ modes) of gravitational waves cannot be treated by using this type of solution.
To advance the study, we therefore need to extend the WWB solution to include both modes \cite{Griffiths:1997}.

\medskip
To construct variations of the WWB solution, we use one of the techniques based on harmonic maps~\cite{Eells}.
As is well known, if a spacetime has two commuting Killing vectors, the Einstein equations are reduced to the so-called Ernst equation, which can be considered from a mathematical viewpoint as the equation that determines a certain kind of harmonic map.
In brief, the harmonic map is defined as the map that gives an extremum of a so-called `energy functional'  
for any maps from a base manifold to a target manifold. 
This mathematical scheme is also used under the name of nonlinear sigma model in theoretical physics. 
The ``energy functional" just corresponds to an ``action functional''.  
In the case of vacuum Einstein equations, gravitational fields can be described compactly with one complex field called Ernst potential, which corresponds to the harmonic map from a virtual three-dimensional Minkowski space to a one-dimensional complex hyperbolic space. 
Because of the nonlinearity of the basic equations to determine such harmonic maps, the explicit construction of the corresponding solutions is a quite difficult task,  though the equation has a simple form.
Nevertheless, the construction of the exact solutions have a long history to overcome the nonlinearity, and several useful methods are now available.

\medskip
Among these methods, here we adopt the most simple one, which is based on the mathematical fact that a composite map of a geodesic curve in a target space and an appropriate harmonic scalar function is a 
harmonic map required~\cite{Eells}. 
In both axially symmetric stationary cases and plane symmetric cases, similar methods have often been used, and many important solutions that describe charged black holes or colliding plane waves have been derived~\cite{book exact solution,book colliding waves}.  
On the other hand, in the cylindrically symmetric case, though solitonic methods~\cite{Belinski:2001ph} or  complex coordinate transformations~\cite{Xanthopoulos:1986bx,Papadopoulos:1990pk,Xanthopoulos:1986ac} have been
 used, there seem to be few works that treat the solutions derived by the simple composite harmonic map. 
In an exceptional work, Halilsoy once used this method to derive some 
cross-polarized solutions from the Einstein-Rosen-type solutions including the WWB solution~\cite{Halilsoy:1988vz}. 
However, it seems that such solutions has not yet been physically  investigated enough, and 
 have been neglected for a long time, 
probably due to 
rather simple expressions of the solutions.

\medskip
Hence as a first step from the point of view mentioned above, 
we construct new variations of the WWB solution including Halilsoy's one by using the method of composite harmonic mapping. 
As described in Sec.III, the solutions considered here have four main parameters $(a,\ c,\ \delta,\ A )$: Using these parameters, we can explore the physical behavior of the gravitational waves corresponding to the solutions. 
Through the systematic analysis of the solutions, we can clarify the nonlinear properties of the gravitational waves, especially by observing the time variation of mutual transformation between the linear mode and the cross mode.

\medskip
As a further application of the generated solutions, it is 
worth noting that we can easily set up collision between multiple cylindrically symmetric waves as a kind of theoretical experiment, owing to the convenience of the solution-generation technique adopted here.
For example, the solution representing collision of one localized wave against a pulselike wave train can 
be constructed easily. 
The solution itself is a rather restricted one, however, its behavior may be  interesting enough.

\medskip
To extend the present analysis to other cylindrically symmetric gravitational systems, the Einstein-Maxwell 
system in four-dimensional spacetime, as a next step, is interesting and appropriate for the application of 
the harmonic mapping method, though the system is more complicated than the Einstein vacuum case. 
In this place, instead of using the same approach as above, we pay attention to a simple, but useful mathematical 
fact as follows: when four physical degrees of freedom of the Einstein-Maxwell equations are consistently reduced 
to two physical degrees of freedom (one is for gravitational fields, and the other is for Maxwell fields), 
these truncated Einstein-Maxwell equations are mathematically equivalent to the vacuum Einstein equations. 
Using this fact, one can show that the set of solutions of the vacuum Einstein equations considered above is 
directly transformed into a subset of solutions of the Einstein-Maxwell equations.
As discussed in Sec VI, this transformation is a rather trivial one from a mathematical viewpoint, 
but once the solutions are interpreted as solutions of Einstein-Maxwell equations, 
the solutions have a different meaning physically, and further may become important in some situations.

\medskip
This paper is organized as follows. 
In Sec.II, we first introduce the basic equations reduced from the vacuum Einstein equations under the assumption of cylindrical symmetry. 
After we briefly explain the composite harmonic mapping method to solve the basic equations, we give a formal expression of the solution to the basic equations using a harmonic function (exactly more saying, a wave function). 
In Sec.III, following the work in Ref.~\cite{Piran}, we define the amplitudes of cylindrical gravitational waves as useful tools for analysis. 
The amplitudes are simply expressed using the harmonic function prepared. 
Next, we proceed to the 
analysis under these preparations. 
In Sec.IV, we generalize Halilsoy's solution, by extending the WWB solution to include the time asymmetry.  
The explicit expression of the obtained solution is summarized. 
In Sec.V, using the solution in Sec.IV, we investigate nonlinear features of the waves. 
This section splits into three parts: in Sec. A, the asymptotic behaviors of the waves are considered, in Sec. B, the behaviors of single waves near a symmetric axis at reflection are examined, 
and in Sec. C, the collisions of gravitational wave pulses are treated.  
In Sec.VI, as further generalizations, we give two other solutions that have explicit expressions. 
On the whole, the analysis is conducted with attention to conversion phenomena between two different polarization modes ($+$ mode and $\times$ mode). 
The final section VI is devoted to summary and discussion, 
where as stated above, physical meaning of mutual transformation between one gravitational mode and 
one electromagnetic mode is also considered.

\section{Basic equations and formal expression of general solutions}

For the spacetimes corresponding to cylindrically symmetric vacuum Einstein equations, we can use the following Kompaneets-Jordan-Ehlers form~\cite{Komaneets-Jordan-Ehlers},
\begin{equation}
ds^2=e^{2\psi}(dz+\omega d\phi)^2+\rho^2 e^{-2\psi}d\phi^2+e^{2(\gamma-\psi)}(-dt^2+d\rho^2),\label{eq:KJE}
\end{equation}
where the functions $\psi$, $\omega$, and $\gamma$ depend on the time coordinate $t$ and the 
radial coordinate $\rho$ only.
Using this metric form, the vacuum Einstein equations are reduced to the following set of equations: 
\begin{eqnarray}
&& \psi,_{tt} - \frac{1}{\rho}\psi,_{\rho}-\psi,_{\rho\rho}
= \frac{e^{4\psi}}{2\rho^2}\left[(\omega,_{t})^2 - (\omega,_{\rho})^2 \right],\label{eq:Ein1} \\
&& \omega,_{tt} + \frac{1}{\rho}\omega,_{\rho}-\omega,_{\rho\rho}
= 4\left( \omega,_{\rho} \psi,_{\rho} - \omega,_{t} \psi,_{t} \right),\label{eq:Ein2} \\
&& \gamma,_{\rho} 
= \rho \left[(\psi,_{t})^2 + (\psi,_{\rho})^2 \right]
+ \frac{e^{4\psi}}{4\rho}\left[(\omega,_{t})^2 + (\omega,_{\rho})^2 \right],\label{eq:Ein3} \\ 
&& \gamma,_{t} 
= 2\rho \psi,_{t} \psi,_{\rho} 
+ \frac{e^{4\psi}}{2\rho} \omega,_{t} \omega,_{\rho}\,. \label{eq:Ein4}
\end{eqnarray}
We further introduce the following quantity ${\bf E}$(called Ernst potential): 
\begin{equation}
{\bf E}= e^{2\psi}+i\Phi\,.
\end{equation}
Here the imaginary part $\Phi$ is defined with $\omega$ through 
\begin{eqnarray}
\Phi{,\,}_{t}=\frac{1}{\rho}e^{4\psi}\omega{,\,}_{\rho}\,,\ \ \ 
\Phi{,\,}_{\rho}=\frac{1}{\rho}e^{4\psi}\omega{,\,}_{t}\,, \label{eq:Phi-def}
\end{eqnarray}
whose integrability is ensured by Eq.~(\ref{eq:Ein2}). 
From the compatibility of the above definition, $\Phi$ satisfies the following equation:
\begin{equation}
\Phi,_{tt} - \frac{1}{\rho}\Phi,_{\rho}-\Phi,_{\rho\rho}
= 4\left( \Phi,_{t}\psi,_{t} - \Phi,_{\rho} \psi,_{\rho} \right)
\end{equation}
The equations for the Ernst potential therefore can be combined into a single equation, 
\begin{equation}
\nabla^2{\bf E}- \frac{2}{{\bf E}+\bar{{\bf E}}}\nabla{\bf E}\cdot \nabla{\bf E}=0\,,\label{eq:Ernst}
\end{equation}
where the operators $\nabla$ and $\nabla^2$ are defined as the gradient and the d'Alembertian 
on a three-dimensional Minkowski spacetime with cylindrical symmetry.
This equation may be considered a cylindrical symmetric version of the Ernst equation originally 
introduced on a stationary axis-symmetric spacetime, so we also call the above equation the Ernst equation. 
The Ernst equation gives harmonic maps from a hypothetical three-dimensional Minkowski space($M^3$) 
to a two-dimensional hyperbolic space($H^2$).
Actually the potential space has the following line element: 
\begin{equation}
ds^2=\frac{2}{({\bf E}+\bar{\bf E})^2}\,d{\bf E}d\bar{\bf E}=\frac{dx^2+dy^2}{y^2}\,,
\end{equation}
which corresponds to the so-callled Poincar$\acute{\rm e}$'s upper half-plane model, 
when $\Phi$ and $e^{2\psi}$ are assigned to the horizontal axis $x$ and the vertical axis $y$ of 
the half plane, respectively. 
As stated in Sec.I, we use the fact that composition of geodesics in a potential space and an appropriate harmonic scalar function is a harmonic map. 
Let us consider composition of $\tau(x):M^3\longrightarrow {\rm R}$ and 
${\bf E}(\tau):{\rm R}\longrightarrow H^{2}$. 
Equation~(\ref{eq:Ernst}) then is written as follows,
\begin{equation}
\left[\frac{d^2{\bf E}}{d\tau^2}- \frac{2}{{\bf E}+\bar{{\bf E}}}\left(\frac{d{\bf E}}{d\tau}\right)^2 \right]
\nabla{\tau}\cdot \nabla{\tau} + \frac{d{\bf E}}{d\tau}\nabla^2 {\tau} \, = \,0 \, . 
\end{equation}
So the composite mapping function ${\bf E}(\tau(x))$ gives a harmonic map when the function $\tau$ and the mapping function ${\bf E}$ satisfy 
the following equations:
\begin{equation}
\nabla^2 {\tau} = 0 \,,\ \ \ \ \ 
\frac{d^2{\bf E}}{d\tau^2}- \frac{2}{{\bf E}+\bar{\bf E}}\left(\frac{d{\bf E}}{d\tau}\right)^2 =0 \,,
\end{equation}
respectively. 
The first equation is just a linear wave equation in a cylindrically symmetric flat spacetime, 
so that general solutions can be represented with Bessel functions formally. 
On the other hand, the nonlinearity is confined into the second equation, which has been reduced to 
a geodesic equation in the hyperbolic space $H^2$. 
For the latter equation the variable $\tau$ plays the role of an affine parameter. 
As a well-known fact, when considering the Poincar$\acute{\rm e}$ half-plane model, 
the geodesics corresponding to the latter equation are represented as semicircles, 
whose centers are on the real axis assigned to $\Phi$, and hence the general 
expression of the geodesics is given as follows, 
\begin{equation}
(\,\Phi\,,\ \,e^{2\psi}\,)
\,=\,\left(\, x_0 + R\,\frac{1-s^2}{1+s^2},\ R\,\frac{2s}{1+s^2}\,\right)\ \ \ 
\,:\,s= e^{2(\tau+\tau_0)}\,,
\end{equation}
where $R$ is a positive constant ({\it i.e.}, radius of the semicircle) and $\tau_0$ is a real constant. 
As a result, once the linear wave function $\tau$ is specified, the corresponding new Ernst potential 
can be obtained directly. 
The above expression still has some gauge redundancy: 
the parameters $x_0$ and $\tau_0$ can take any real number, independently. 
Using this arbitrariness, one can show that after setting $x_0$ to $R$ and replacing $s$ with $s/A$, the Ernst potential has an overall factor $2RA$. 
This factor can be absorbed by scaling the coordinates 
$z$ and $\phi$, and renormalizing the metric 
coefficient $\gamma$. 
Finally, as a general expression, ${\bf E}$ is reduced to the following simple form: 
\begin{equation}
{\bf E} \,=\,\frac{s}{s^2+A^2} \,+\, i\, \frac{A}{s^2+A^2}\,. \label{eq:Ehles}
\end{equation}
It is noted here that if we take $s^{-1}$ as the seed of the Ernst potential, the same expression is also 
derived by the Ehlers transformation. 
For later convenience, some useful formulas are presented. 
From Eq.~(\ref{eq:Ehles}), each part of the Ernst 
potentials is given, 
\begin{eqnarray}
e^{2\psi} = \frac{1}{e^{-2\tau} + A^2 e^{2\tau}}\,,\ \ 
\Phi = \frac{Ae^{2\tau}}{e^{-2\tau} + A^2 e^{2\tau}}\,, \label{eq:psi}
\end{eqnarray}
with $s=e^{-2\tau}$, respectively.
From Eq.~(\ref{eq:Phi-def}), the metric component $\omega$ can be expressed as a contour integral 
along an appropriate path,
\begin{eqnarray}
\omega = 4A \int\rho[\,\tau,_{t}\,d\rho + \tau,_{\rho}\,dt \,]\,. \label{eq:omega2}
\end{eqnarray}
The following formulas are also useful:
\begin{eqnarray}
\tatevec{ \psi,_{t} }{ \psi,_{\rho} }
=
\frac{e^{-2\tau} - A^2 e^{2\tau} }{ e^{-2\tau} + A^2 e^{2\tau} }
\tatevec{ \tau,_{t} }{\tau,_{\rho} }\,,
\ \ \ 
\frac{e^{2\psi}}{2\rho}\tatevec{ \omega,_{t} }{ \omega,_{\rho} }
=
\frac{4A }{ e^{-2\tau} + A^2 e^{2\tau} }
\tatevec{ \tau,_{\rho} }{\tau,_{t} }. 
\end{eqnarray}
Using the above formulas, we can easily derive the equations that determine $\gamma$ as follows, 
\begin{eqnarray}
\gamma,_{t} = 2\rho \tau,_{t}\tau,_{\rho} \,,\ \ \ 
\gamma,_{\rho} = \rho \left[ (\tau,_{t})^2 + (\tau,_{\rho})^2 \right] \,. \label{eq:gamma1}
\end{eqnarray}
The metric function $\gamma$ is the so-called C-energy, which is very useful for 
the later analysis, which plays a role of ``gravitational energy"~\cite{Thorn}. 
It is hence shown from the above equations that this {\rm C}-energy has no dependence on the parameter $A$. 
In fact, this is ensured by the fact that the right-hand sides of Eqs.~(\ref{eq:Ein3}) and (\ref{eq:Ein4}) are invariants 
under the isometry of the potential space.

\section{Amplitudes}
When one considers the propagation of cylindrically symmetric gravitational waves with two polarization modes, it is of great convenience to use the definitions for the amplitudes of the nonlinear waves in Ref.~\cite{Piran},  which are defined as follows: 
The ingoing amplitudes $(A_+,A_\times)$ and outgoing amplitudes $(B_+,B_\times)$ are given, respectively,  by 
\begin{eqnarray}
&&A_+=2\psi_{, v},\label{eq:Ap}\\
&&B_+=2\psi_{, u},\label{eq:Bp}\\
&&A_\times=\frac{e^{2\psi}\omega_{, v}}{\rho},\label{eq:Ac}\\
&&B_\times=\frac{e^{2\psi}\omega_{, u}}{\rho},\label{eq:Bc}
\end{eqnarray}
where the ingoing and outgoing null coordinates $u$ and $v$ are defined by $ u=(t-\rho)/2$ and $ v=(t+\rho)/2$, respectively.
The indices $+$ and $\times$ denote the quantities associated with the respective polarizations. 
Then, the vacuum Einstein equations~(\ref{eq:Ein1})-(\ref{eq:Ein4}) can be written only in terms of these quantities, actually; the nonlinear differential equations (\ref{eq:Ein1}) and (\ref{eq:Ein2}) for the functions $\psi$ and $\omega$ are replaced by
\begin{eqnarray}
&&A_{+, u}=\frac{A_+-B_+}{2\rho}+A_\times B_\times,\label{eq:Apu}\\
&&B_{+, v}=\frac{A_+-B_+}{2\rho}+A_\times B_\times,\label{eq:Bpv}\\
&&A_{\times, u}=\frac{A_\times+B_\times}{2\rho}-A_+ B_\times,\label{eq:Acu}\\
&&B_{\times, v}=-\frac{A_\times+B_\times}{2\rho}+A_\times B_+,\label{eq:Bcv}
\end{eqnarray}
and a couple of the equations (\ref{eq:Ein3}) and (\ref{eq:Ein4}) that determine the function $\gamma$ can be rewritten in terms of the amplitudes as    
\begin{eqnarray}
&&\gamma_{,\rho}=\frac{\rho}{8}\left(A_+^2+B_+^2+A_\times^2+B_\times^2\right),\label{eq:gammarho}\\
&&\gamma_{,t}=\frac{\rho}{8}\left(A_+^2-B_+^2+A_\times^2-B_\times^2\right).
\end{eqnarray}
In particular, if the metric takes a diagonal form, {\it i.e.}, $A_\times=B_\times=0$, Eqs.~(\ref{eq:Apu})-(\ref{eq:Bcv}) are reduced to the set of  the linear equations,
\begin{eqnarray}
&&A_{+, u}=\frac{A_+-B_+}{2\rho},\\
&&B_{+, v}=\frac{A_+-B_+}{2\rho},
\end{eqnarray}
which is another form of the wave equation ${\bm \nabla}^2\psi=0$.
 
From Eq.(\ref{eq:gammarho}), $\gamma_{,\rho}$ can be naturally  interpreted as the total energy density of cylindrical waves, which we denote by ${\cal E}$ in what follows.  
The energy densities assigned to the $+$ mode and $\times$ mode, respectively, as
\begin{eqnarray}
{\cal E}_+:&=&\frac{\rho}{8}\left(A_+^2+B_+^2\right), \label{eq:Ep}\\
{\cal E}_\times:&=&\frac{\rho}{8}\left(A_\times^2+B_\times^2\right).\label{eq:Ec}
\end{eqnarray}
This C-energy density $\gamma_{,\rho}$ is locally measurable for an observer along the world line $\rho$=const.. 
According to Thorn~\cite{Thorn}, the total energy per unit length 
of $z$ 
contained within the radius $\rho_0$ (: constant) at a certain time $t$ is defined 
with ${\cal E}$ as
\begin{eqnarray}
E(t,\rho_0)=\int_0^{\rho_0}{\cal E}d\rho=\gamma(t,\rho_0)-\gamma(t,0).
\end{eqnarray}
Similarly, the total energies assigned to the $+$ and $\times$ modes can be defined, respectively, as
\begin{eqnarray}
E_+(t,\rho_0)&=&\int_0^{\rho_0}{\cal E}_+d\rho,\\
E_\times(t,\rho_0)&=&\int_0^{\rho_0}{\cal E}_\times d\rho.
\end{eqnarray} 

Note that for the expression obtained in the previous section, the wave amplitudes can be generally written as
\begin{eqnarray}
A_+&=&2e^{2\psi}(e^{-2\tau}-A^2e^{2\tau})(\tau_{,t}+\tau_{,\rho}),\\
B_+&=&2e^{2\psi}(e^{-2\tau}-A^2e^{2\tau})(\tau_{,t}-\tau_{,\rho}),\\
A_\times&=&4Ae^{2\psi}(\tau_{,t}+\tau_{,\rho}),\\
B_\times&=&-4Ae^{2\psi}(\tau_{,t}-\tau_{,\rho}).
\end{eqnarray}
Therefore, the ratios of the $\times$ mode to the $+$ mode are given by
\begin{eqnarray}
\frac{B_\times^2}{B_+^2}=\frac{4A^2}{(e^{-2\tau}-A^2e^{2\tau})^2},\quad \frac{A_\times^2}{A_+^2}=\frac{4A^2}{(e^{-2\tau}-A^2e^{2\tau})^2}.\label{eq:ratio}
\end{eqnarray}

It immediately turns out from the above equations that the ratios ${\cal E}_+/{\cal E}$ and ${\cal E}_\times/{\cal E}$ can be expressed, respectively, as
\begin{eqnarray}
&&\frac{\cal E_+}{\cal E}=\left(\frac{e^{-2\tau}-A^2e^{2\tau}}{e^{-2\tau}+A^2e^{2\tau}}\right)^2,\\
&&\frac{\cal E_\times}{\cal E}=\left(\frac{2A}{e^{-2\tau}+A^2e^{2\tau}}\right)^2,
\end{eqnarray}
and the necessary and sufficient condition for ${\cal E}_\times>{\cal E}_+$ is 
\begin{eqnarray}
\frac{1}{2}\ln\frac{\sqrt{2}-1}{A}<\tau< \frac{1}{2}\ln\frac{\sqrt{2}+1}{A}, \label{eq:condition}
\end{eqnarray}
which is of great use for the later analysis of the exact solutions.

\section{Generalization of the Halilsoy solution}\label{sec:sol1}
In this section,  we generalize the Halilsoy pulse solution in Ref.~\cite{Halilsoy:1988vz}, which was obtained by performing the harmonic mapping method for the WWB pulse. 
To this end, first, we generalize the WWB  solution 
to have the additional parameter that presents the extent of time asymmetry. 
Next,  by regarding the generalized WWB solution as a seed, we can present the generalization of the Halilsoy pulse solution with the additional parameter.

Let us recall that for the $0$th order Bessel function $J_0(x)$, the following integral formula holds
\begin{eqnarray}
\int_0^\infty e^{-\alpha k}J_0(k\rho)dk=\frac{1}{\sqrt{\rho^2+\alpha^2}},
\end{eqnarray}
where $\alpha$ is a constant.
If one considers the constant  $\alpha$ to be a complex number and puts $\alpha=a-it$ ($t$ and $a$ are a time coordinate and  a positive constant, respectively),  one can obtain the following formal expression 
\begin{eqnarray}
\int_0^\infty e^{-k(a-it)}J_0(k\rho)dk=\frac{1}{\sqrt{\rho^2+(a-it)^2}}.
\end{eqnarray}
The real and imaginary parts of the above equation are written, respectively, as 
\begin{eqnarray}
&&\int_0^\infty e^{-ka}J_0(k\rho)\cos(kt)dk\nonumber \\
&&\hspace{2cm} =\frac{1}{\sqrt{2}}\sqrt{\frac{\sqrt{(t^2-\rho^2-a^2)^2+4a^2t^2}-(t^2-\rho^2-a^2)}{(t^2-\rho^2-a^2)^2+4a^2t^2}},\label{eq:v0}\\
&&\int^\infty_0e^{-ka}J_0(k\rho)\sin(kt)dk\nonumber\\
           &&\hspace{2cm} =\frac{\sqrt{2}at}{\sqrt{(t^2-\rho^2-a^2)^2+4a^2t^2} \sqrt{\sqrt{(t^2-\rho^2-a^2)^2+4a^2t^2}-(t^2-\rho^2-a^2)}}.
\end{eqnarray}
In terms of the new coordinates $(x,y)$ defined as
\begin{eqnarray}
\rho=a\sqrt{(x^2+1)(y^2-1)},\quad t=axy, \label{eq:trans}
\end{eqnarray}
the real part and the imaginary part multiplied by the constants $c$ and $c/\sqrt{2}$, respectively,  can be written in the considerably simple forms
\begin{eqnarray}
&&\tau_{\rm even}=c\int_0^\infty e^{-ka}J_0(k\rho)\cos(kt)dk
=\frac{c}{a}\frac{y}{x^2+y^2},\\
&&\tau_{\rm odd}=\frac{c}{\sqrt{2}}\int_0^\infty e^{-ka}J_0(k\rho)\sin(kt)dk
=\frac{c}{a}\frac{x}{x^2+y^2}.
\end{eqnarray}

Moreover, superposing the two solutions to the wave equation, we can obtain the more general solution as follows,
\begin{eqnarray}
\tau
  &=&c(\tau_{\rm even}\cos\delta+\tau_{\rm odd}\sin\delta)\\
  &=&\frac{c}{a}\frac{y\cos\delta+x\sin\delta}{x^2+y^2}, \label{eq:v}
\end{eqnarray}
where $\delta$ is a real constant.
Substituting the above $\tau$  for the right-hand side of Eq.~(\ref{eq:omega2}), we can integrate $\omega$ to find
\begin{eqnarray}
\omega&=&\omega_{\rm even}\cos\delta+\omega_{\rm odd}\sin\delta, \label{eq:omega}
\end{eqnarray}
where $\omega_{\rm even}$ and $\omega_{\rm odd}$ are obtained by the substitutions of $\tau_{\rm even}$ and $\tau_{\rm odd}$, respectively,for the right-hand side of Eq.~(\ref{eq:omega2}), to be determined, up to a constant, as
\begin{eqnarray}
\omega_{\rm even}&=&-4Ac\frac{x(y^2-1)}{x^2+y^2},\\
\omega_{\rm odd}&=&-4Ac\frac{y(x^2+1)}{x^2+y^2}.
\end{eqnarray}
Also, replacing $\tau$ in Eq.~(\ref{eq:gamma1}) with Eq.~(\ref{eq:v}), we can integrate $\gamma$ to give
\begin{eqnarray}
\gamma=\gamma_{\rm even}\cos^2\delta+\gamma_{\rm odd}\sin^2\delta+\gamma_{\rm cross}\sin2\delta+\frac{c^2}{4a^2}, \label{eq:gamma}
\end{eqnarray}
where
\begin{eqnarray}
\gamma_{\rm even}&=&\frac{c^2 \left(x^2+1\right) \left[-x^6+x^4 \left(1-4 y^2\right)+3 x^2 y^2 \left(y^2-2\right)-2 y^6+y^4\right]}{4 a^2 \left(x^2+y^2\right)^4}, \label{eq:gammae}\\
\gamma_{\rm odd}&=&-\frac{c^2 \left(x^2+1\right) \left[x^6+x^4 \left(2 y^2+1\right)+x^2 \left(9 y^4-6 y^2\right)+y^4\right]}{4 a^2 \left(x^2+y^2\right)^4},\label{eq:gammao}\\
\gamma_{\rm cross}&=&\frac{4c^2 xy\left(x^2+1\right) \left(y^2-1\right) \left(x^2-y^2\right)}{4 a^2 \left(x^2+y^2\right)^4}.\label{eq:gammac}
\end{eqnarray}
Thus, regarding the generalized WWB pulse solution as the seed of the harmonic mapping, we have obtained the generalization of the Halilsoy solution given by the three functions  $(\tau,\omega,\gamma)$,  which depends on only $t$ and $\rho$, and has the four parameters $(a,\ c,\ \delta,\ A )$.   
The first two parameters $a$ and $c$ are already held by the WWB solution, and are related to the width and amplitude of the wave, respectively. 
The third parameter $\delta$ is first introduced here to show the extent of time asymmetry of the solutions. 
The fourth parameter $A$ implies the measurement of the extent of nonlinearity, in particular, the case $A=0$ corresponds to the Einstein-Rosen-type of linear waves. 
Note that  the case $\delta=0$ [namely, the solution expressed by only $(\tau_{\rm even},\omega_{\rm even},\gamma_{\rm even})$] corresponds to the Halilsoy solution~\cite{weber:1957,bonnor:1957}.
In the following section, by varying these parameters, we see the physical behavior of the gravitational waves that the solutions describe.

\section{Analysis}
In this section, we  study the nonlinear properties of the gravitational waves through the analysis of the exact solutions obtained in the previous section, especially by observing the time variation of the mutual transformation between the $+$ mode and $\times$ mode.
First of all, we focus on the asymptotic behaviors of gravitational pulse waves in the neighborhood of the spacetime boundaries, the axis $\rho=0$ with $t=$constant, spacelike infinity $\rho\to \infty$ with $t$ constant, future null infinity $v\to \infty$ with $u$ constant and past null infinity $u\to -\infty$ with $v$ constant. 
Next, we also consider how much the nonlinear interaction converts the $+$ mode ($\times$ mode) pulse wave to the $\times$ mode ($+$ mode) pulse wave (such a nonlinear effect is often called the gravitational Faraday effect) at the reflection of the gravitational pulse waves at the axis.
Finally, we numerically study how the mode conversion occurs when two single pulses, an outgoing gravitational pulse and an ingoing pulse, collide, and also when a single outgoing pulse and multi-ingoing pulses collide.

\subsection{Asymptotics}

Near the axis $\rho=0$ $(t={\rm const.})$,  the metric is approximated by
\begin{eqnarray}
ds^2&\simeq& e^{\frac{2c(a\cos \delta+t \sin \delta)}{a^2+t^2}} \left(1+A^2 e^{\frac{4c(a\cos\delta+t\sin\delta)}{a^2+t^2}}\right)^{-1}\left(dz-4Ac\sin\delta d\phi\right)^2\nonumber\\
&&+\rho^2e^{-\frac{2c(a\cos \delta+t \sin \delta)}{a^2+t^2}} \left(1+A^2 e^{\frac{4c(a\cos\delta+t\sin\delta)}{a^2+t^2}}\right)d\phi^2\\ \nonumber
&&+e^{-\frac{2c(a\cos \delta+t \sin \delta)}{a^2+t^2}} \left(1+A^2 e^{\frac{4c(a\cos\delta+t\sin\delta)}{a^2+t^2}}\right)(-dt^2+d\rho^2).
\end{eqnarray}
From this asymptotic form, it is straightforward to show the absence of the deficit angle $\Delta\phi$, namely, 
\begin{eqnarray}
\Delta\phi&:=&2\pi-\lim_{\rho \to 0}\frac{\int^{2\pi}_0\sqrt{g_{\phi\phi}}d\phi}{\int^\rho_0\sqrt{g_{\rho\rho}}d\rho}\\
               &=&0.
\end{eqnarray}
This means that if one chooses the periodicity of the angular coordinate $\phi$ to be $0\le \phi<2\pi$, no deficit angle is present on the axis. 
Near the axis $\rho=0$, the energy densities corresponding to the $+$ mode and $\times$ mode waves behave, respectively, as 
\begin{eqnarray}
&&\frac{\rho}{8}A_+^2\simeq \frac{\rho}{8}B_+^2\simeq \frac{c^2\left(1-A^2e^{\frac{2c(2a\cos\delta+t\sin\delta)}{a^2+t^2}}\right)^2(-4at\cos\delta+(a^2-t^2)\sin\delta)^2}{8\left(1+A^2e^{\frac{2c(2a\cos\delta+t\sin\delta)}{a^2+t^2}}\right)^2(a^2+t^2)^4}\rho,\\
&&\frac{\rho}{8}A_\times^2\simeq \frac{\rho}{8}B_\times^2\simeq \frac{c^2A^2e^{\frac{2c(2a\cos\delta+t\sin\delta)}{a^2+t^2}}(-4at\cos\delta+(a^2-t^2)\sin\delta)^2}{2\left(1+A^2e^{\frac{2c(2a\cos\delta+t\sin\delta)}{a^2+t^2}}\right)^2(a^2+t^2)^4}\rho.
\end{eqnarray}

\medskip
At spacelike infinity $\rho\to \infty$ $(t={\rm const.})$, we have the asymptotic form of the metric
\begin{eqnarray}
ds^2\simeq \frac{1}{1+A^2}dz^2+\rho^2(1+A^2)d\phi^2+(1+A^2)e^{-\frac{c^2}{2a^2}}(-dt^2+d\rho^2).
\end{eqnarray}
 It tunrs out that the spacetime is locally Minkowski spacetime since this metric can be written as
 \begin{eqnarray}
ds^2\simeq (dz')^2+\rho^{\prime 2}(d\phi')^2-(dt')^2+(d\rho')^2,
\end{eqnarray}
where we have defined the new coordinates $(z',\phi',t',\rho')$ such that
\begin{eqnarray}
z'=\frac{z}{\sqrt{1+A^2}},\quad \phi'=\sqrt{1+A^2}\ e^{\frac{c^2}{4a^2}}\phi,\quad  t'=\sqrt{1+A^2}\ e^{-\frac{c^2}{4a^2}}t,\quad  \rho'=\sqrt{1+A^2}\ e^{-\frac{c^2}{4a^2}}\rho.
\end{eqnarray}
However, the periodicity of $\phi'$ cannot be $2\pi$ as long as one assumes the periodicity of $\phi$ is $2\pi$. 
In fact, at infinity, the “deficit angle" can be computed as
\begin{eqnarray}
\Delta\phi&:=&2\pi-\lim_{\rho \to \infty}\frac{\int^{2\pi}_0\sqrt{g_{\phi\phi}}d\phi}{\int^\rho_0\sqrt{g_{\rho\rho}}d\rho}\\
               &=&2\pi(1-e^{-\frac{c^2}{4a^2}}).
\end{eqnarray} 
It is obvious that the presence of this deficit angle is due to the energy of the gravitational pulse waves. 
At infinity,  the energy densities behave as
\begin{eqnarray}
&&\frac{\rho}{8}A_+^2\simeq \frac{\rho}{8}B_+^2\simeq \frac{c^2(1-A^2)^2\cos^2\delta}{2(1+A^2)^2\rho^3},\\
&&\frac{\rho}{8}A_\times^2\simeq \frac{\rho}{8}B_\times^2\simeq \frac{2c^2A^2\cos^2\delta}{(1+A^2)^2\rho^3},
\end{eqnarray}
which become swiftly smaller as $\rho$ becomes larger.

\medskip
Let us consider the asymptotic behaviors of the pulse waves at timelike infinity $t\to\pm \infty$ ($\rho=$const.). 
At $t\to\pm\infty$, we have 
\begin{eqnarray}
ds^2\simeq \frac{1}{1+A^2}\left(dz-4Ac\sin\delta d\phi\right)^2+\rho^2(1+A^2)d\phi^2+(1+A^2)e^{-\frac{c^2}{2a^2}}(-dt^2+d\rho^2),
\end{eqnarray}
which shows that at late (early) time, the spacetime approaches Minkowski spacetime. Actually, in terms of the new coordinates $(z'',\phi'',t'',\rho'')$
\begin{eqnarray}
z''=\frac{z-4A\sin(\delta)\  \phi}{\sqrt{1+A^2}},\quad \phi''=\sqrt{1+A^2}\ e^{\frac{c^2}{4a^2}}\phi,\quad  t''=\sqrt{1+A^2}\ e^{-\frac{c^2}{4a^2}}t,\quad  \rho''=\sqrt{1+A^2}\ e^{-\frac{c^2}{4a^2}}\rho,
\end{eqnarray}
the asymptotic form of the metric turns out to be written as
\begin{eqnarray}
ds^2\simeq (dz'')^2+\rho^{\prime\prime 2}(d\phi'')^2-(dt'')^2+(d\rho'')^2.
\end{eqnarray}
Since at both late and early times, the energy densities behave as
\begin{eqnarray}
&&\frac{\rho}{8}A_+^2\simeq \frac{\rho}{8}B_+^2\simeq \frac{\left(A^2-1\right)^2 c^2 \rho  \sin ^2\delta }{8 \left(A^2+1\right)^2 t^4},\\
&&\frac{\rho}{8}A_\times^2\simeq \frac{\rho}{8}B_\times^2\simeq \frac{A^2 c^2 \rho  \sin ^2\delta }{2\left(A^2+1\right)^2 t^4},
\end{eqnarray}
the tails of the gravitational waves become smaller and smaller.

\medskip
At future null infinity $v\to\infty$ ($u=$const.), the metric behaves as
\begin{eqnarray}
ds^2&\simeq&\frac{1}{1+A^2}\left[dz-2Ac\frac{(\sqrt{F(-u)}F(u)\cos\delta-\sqrt{F(u)}F(-u)\sin\delta )\sqrt{v}}{a\sqrt{a^2+u^2}}d\phi \right]^2\nonumber\\
       &+&\rho^2(1+A^2)d\phi^2+(1+A^2)e^{F_+}(-dt^2+d\rho^2),
\end{eqnarray}
where $F(x):=\sqrt{a^2+4x^2}+2x$ and 
\begin{eqnarray}
F_+=\frac{c^2[a^2(a^2-4u^2)\cos2\delta-4a^3u\sin2\delta-2(a^2+4u^2)(a^2+2uF(u))]}{8(a^2+4u^2)^2a^2}.
\end{eqnarray}
The energy densities of the outgoing waves 
corresponding to
the $+$ and $\times$ modes behave as, respectively, 
\begin{eqnarray}
\frac{\rho}{4}B_+^2&\simeq& \frac{c^2(1-A^2)^2[-2(a^2+2uF(-u))\cos\delta+a(F(-u)-2u)\sin\delta]^2}{16(1+A^2)^2(a^2+4u^2)^3F(-u)},\\
\frac{\rho}{4}B_\times^2&\simeq& \frac{c^2A^2[-2(a^2+2uF(-u))\cos\delta+a(F(-u)-2u)\sin\delta]^2}{4(1+A^2)^2(a^2+4u^2)^3F(-u)}.
\end{eqnarray}

At past null infinity $u\to -\infty$  ($v=$const.), the metric behaves as
\begin{eqnarray}
ds^2&\simeq&\frac{1}{1+A^2}\left[dz+2Ac\frac{(\sqrt{F(-v)}F(v)\cos\delta-\sqrt{F(v)}F(-v)\sin\delta )\sqrt{-u}}{a\sqrt{a^2+v^2}}d\phi \right]^2\nonumber\\
       &+&\rho^2(1+A^2)d\phi^2+(1+A^2)e^{F_-}(-dt^2+d\rho^2),
\end{eqnarray}
where
\begin{eqnarray}
F_-=\frac{c^2[a^2(a^2-4v^2)\cos2\delta-4a^3v\sin2\delta-2(a^2+4v^2)(a^2-2vF(-v))]}{8(a^2+4v^2)^2a^2}.
\end{eqnarray}
Similarly, 
the energy densities of the ingoing waves 
corresponding to 
the $+$ and $\times$ modes behave as, respectively,
\begin{eqnarray}
\frac{\rho}{4}A_+^2&\simeq& \frac{c^2(1-A^2)^2[2(a^2-2vF(v))\cos\delta+a(F(v)+2v)\sin\delta]^2}{16(1+A^2)^2(a^2+4v^2)^3F(v)},\\
\frac{\rho}{4}A_\times^2&\simeq& \frac{c^2A^2[2(a^2-2uF(v))\cos\delta+a(F(v)-2v)\sin\delta]^2}{4(1+A^2)^2(a^2+4v^2)^3F(v)}.
\end{eqnarray}
Therefore,  it turns out that the ratios of the $\times$ mode to the $+$ mode at future and past null infinities are, respectively, 
\begin{eqnarray}
\frac{B_\times^2}{B_+^2}=\frac{4A^2}{(1-A^2)^2},\quad \frac{A_\times^2}{A_+^2}=\frac{4A^2}{(1-A^2)^2}.\label{eq:ratio2}
\end{eqnarray}
This means that these two ratios are exactly equal. 
It is worth noting that  this is a  general fact because the result can be derived directly from Eq.~(\ref{eq:ratio}) 
if the harmonic function $\tau$ vanishes at future and past null infinities, respectively, as
\begin{eqnarray}
\tau&\simeq&{\cal O}((-u)^{-\frac{1}{2}}),\\
\tau&=&{\cal O}(v^{-\frac{1}{2}}).
\end{eqnarray}

\subsection{Reflection}
By observing the behaviors of the total energy density $\cal E$, the $+$ mode portion $\cal E_+$ and the $\times$ mode portion $\cal E_{\times}$ around the axis, we show, here, some interesting behaviors of the gravitational waves that embody genuine nonlinearity of 
gravitational interaction. 

\begin{figure}[h]
  \begin{tabular}{ccc}
 \begin{minipage}[t]{0.3\hsize}
 \centering
\includegraphics[width=5.7cm]{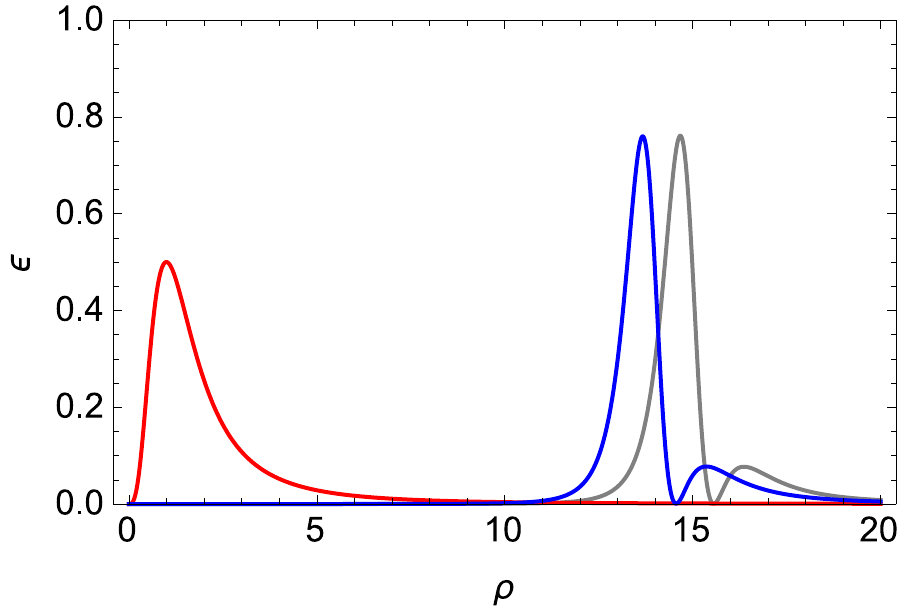}
 \end{minipage} &\ \ \ \ \ \ 
 
 \begin{minipage}[t]{0.3\hsize}
 \centering
\includegraphics[width=5.7cm]{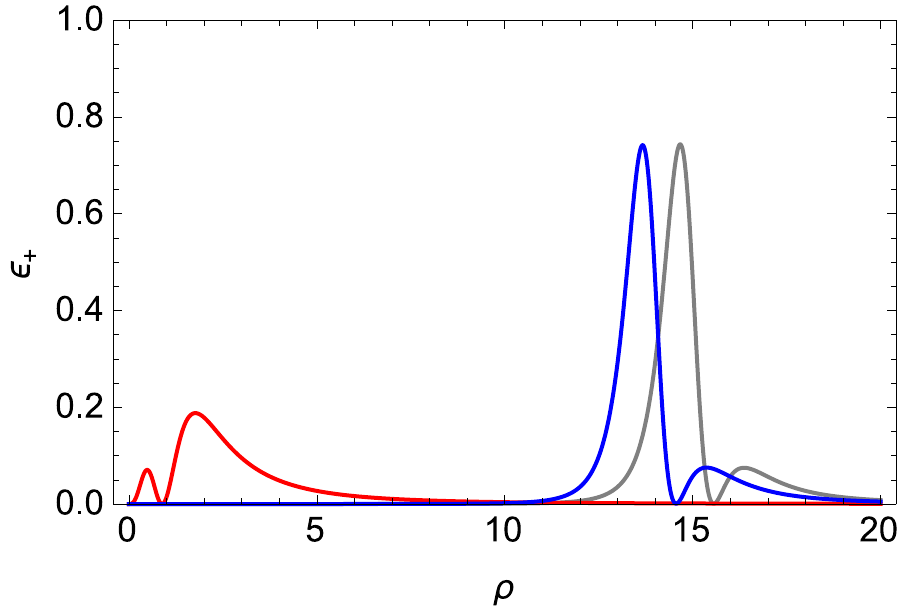}
 \end{minipage} &\ \ \ \ \ \ 
 
  \begin{minipage}[t]{0.3\hsize}
 \centering
\includegraphics[width=5.7cm]{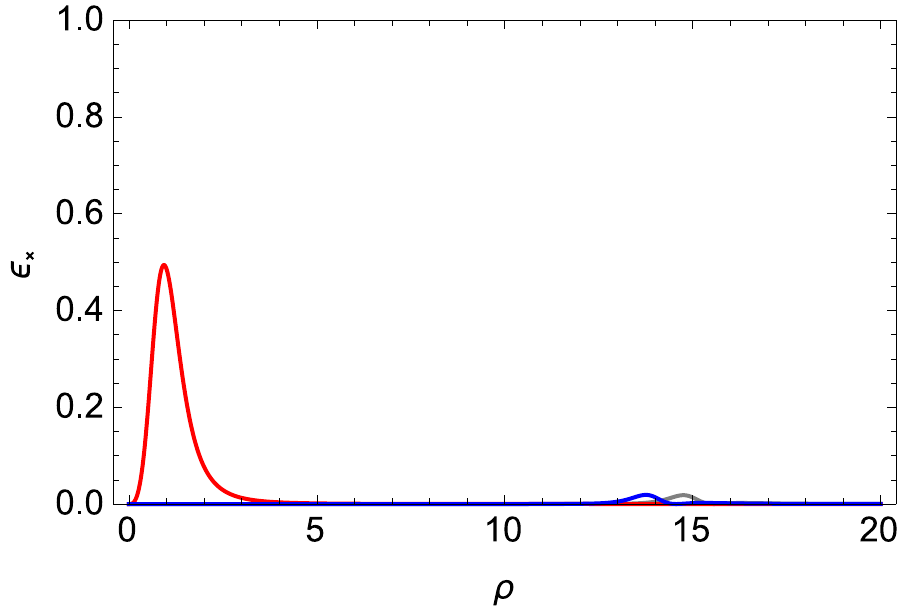}
 \end{minipage} 

 \end{tabular}
\caption{Time dependence of the energy densities ${\cal E}$, ${\cal E}_+$  and ${\cal E}_\times$ for  $(a,c,A,\delta)=(1,2,0.05,0)$. Each figure displays the snapshots of $\cal E$, $\cal E_+$, and $\cal E_\times$  in the order from left to right, where the gray, red, and blue graphs show the pulses corresponding to $t=-15,0,14$, respectively.}
\label{fig:reflection1}
\end{figure}

\medskip
As a characteristic parameter set, we take $(a,c,A,\delta)=(1,2,0.05,0)$, for example. 
The left, middle, and right graphs in Fig.~\ref{fig:reflection1} display the time evolution of $\cal E$, $\cal E_+$,  
and $\cal E_\times$, respectively, where gray, red, and blue ones in each graph show the instantaneous 
figures corresponding to $t=-15,0,14$, respectively.  
From these behaviors of the energy densities, we may consider that the gravitational pulse is a regular wave packet 
that first comes into the region near the symmetric axis from past null infinity and leaves the axis after reflection 
for the future null infinity. 
It should be especially remarked from Fig.\ref{fig:reflection1} that the latter two graphs are different from the 
first graph, which shows the behavior of total energy density, 
 namely, the $+$ mode is dominant away from the axis of symmetry, whereas the $\times$ mode can be comparable to the $+$ mode near the axis.

\begin{figure}[h]
  \begin{tabular}{ccc}

 \begin{minipage}[t]{0.3\hsize}
 \centering
\includegraphics[width=5.7cm]{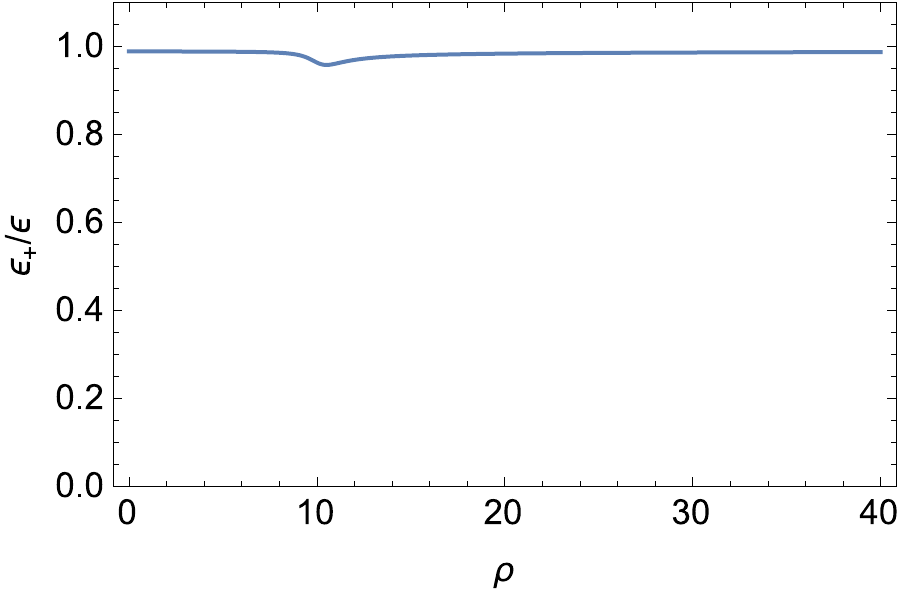}
 \end{minipage} &\ \ \ \ \ \ 
 
  \begin{minipage}[t]{0.3\hsize}
 \centering
\includegraphics[width=5.7cm]{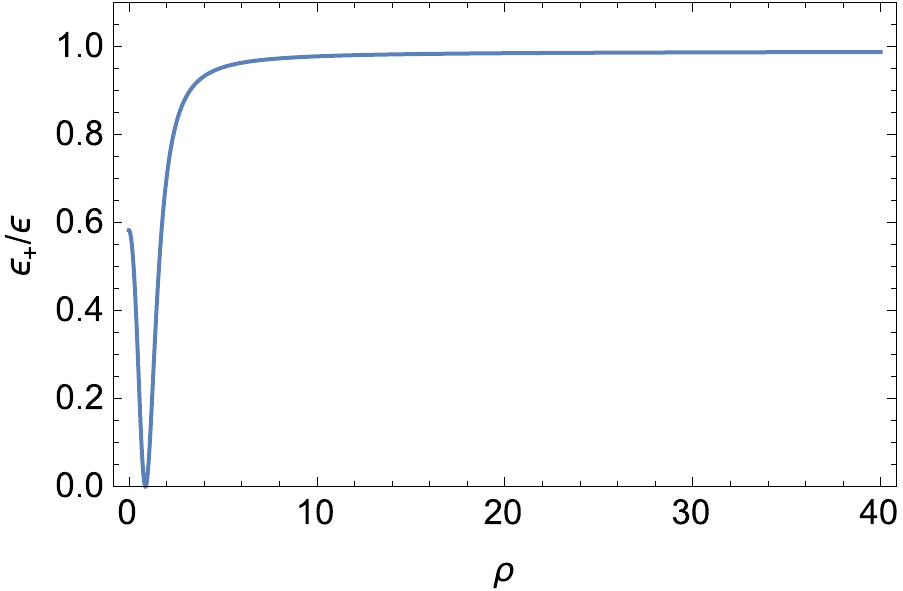}
 \end{minipage}&\ \ \ \ \ \

 \begin{minipage}[t]{0.3\hsize}
 \centering
\includegraphics[width=5.7cm]{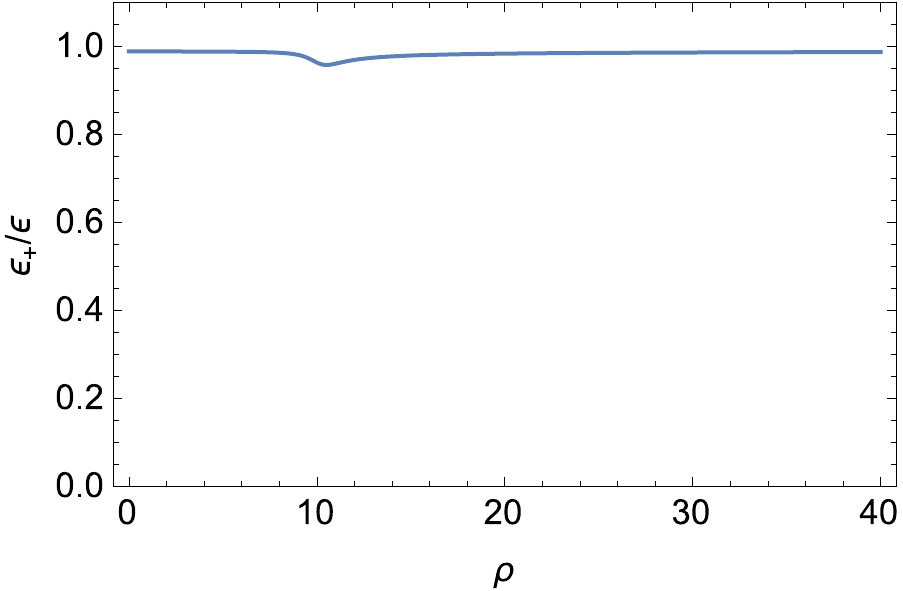}
 \end{minipage}

 \end{tabular}
 \caption{Time dependence of the ratio $\cal E_+/\cal E$ for  $(a,c,A,\delta)=(1,2,0.05,0)$. In the figures from left to right, the graphs display the ratio at 
 $t=-10,0,10$}
 \label{fig:Reflection-ratio1}
\end{figure}

\medskip
The three graphs in Fig.~\ref{fig:Reflection-ratio1} show the time dependence of the ratio of  the energy density of the $+$ mode wave to the total energy density,  $\cal E_+/\cal E$, at each time of $t=-10,0,10$. It can be found that when the $+$ mode wave that comes in from infinity  reflects at the axis, the pulse with the $\times$ mode is generated just for a moment but it soon vanishes, 
and the pulse 
of almost the +mode 
goes out to the infinity. We have numerically confirmed that in each case of $A\ll 1$, the behaviors of a pulse wave are considerably similar.

\begin{figure}[h]

  \begin{tabular}{ccc}
 \begin{minipage}[t]{0.45\hsize}
 \centering
\includegraphics[width=8cm]{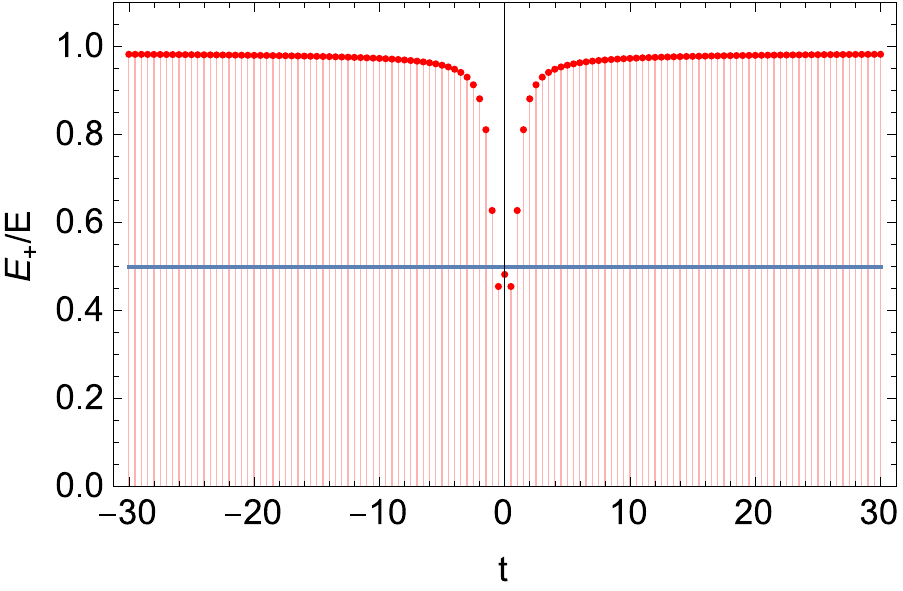}
 \end{minipage} &\ \ \ \ \ \ 
 
 \begin{minipage}[t]{0.45\hsize}
 \centering
\includegraphics[width=8cm]{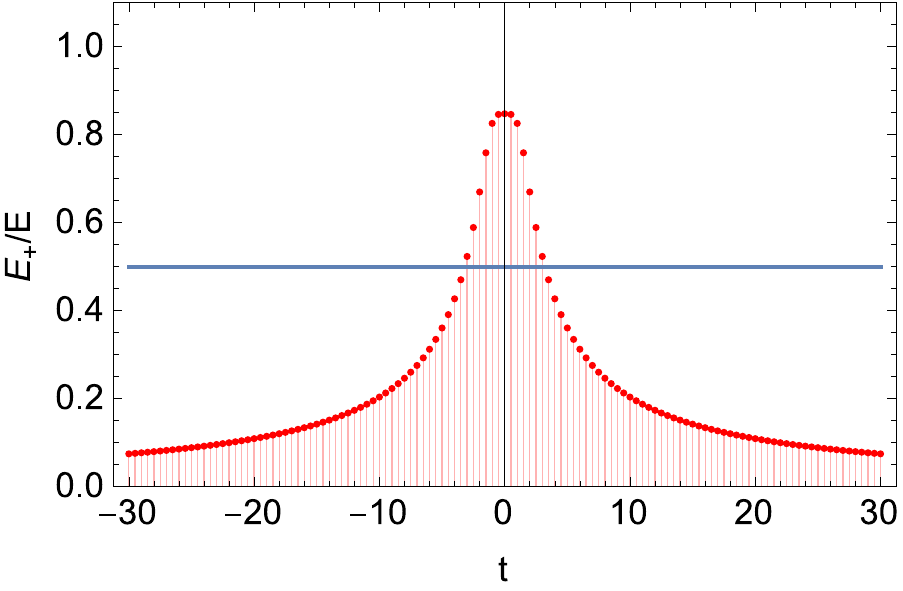}
 \end{minipage} &\ \ \ \ \ \

 \end{tabular}
\caption{
The left and right figures display the time dependence of the ratio $E_+(t,1000)/E(t,1000)$
 for $(a,c,A,\delta)=(1,2,0.05,0)$ and $(a,c,A,\delta)=(1,2,1,0)$, respectively. }
\label{fig:integ1}
\end{figure}

Next, by observing the time dependence of the ratio $E_+(t,\rho_0)/E(t,\rho_0)$, 
let us see how the C-energy is 
converted btween the +mode and $\times$ mode. 
The left graph in Fig.~\ref{fig:integ1} shows the time dependence of $E_+(t,1000)/E(t,1000)$  for $-30\le t\le 30$. 
It can be seen from this graph that only when the incident pulse with the $+$ mode that comes in from infinity reflects at the axis around $t=0$, a large amount of energy of the $+$ mode is transformed into the energy of the $\times$ mode.

\begin{figure}[h]
  \begin{tabular}{ccc}
 \begin{minipage}[t]{0.3\hsize}
 \centering
\includegraphics[width=5.7cm]{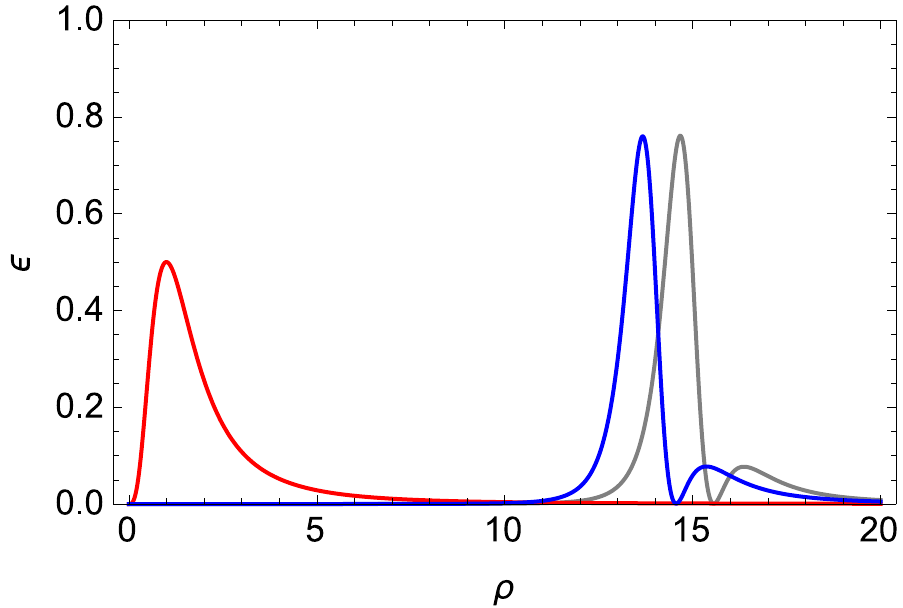}
 \end{minipage} &\ \ \ \ \ \ 
 
 \begin{minipage}[t]{0.3\hsize}
 \centering
\includegraphics[width=5.7cm]{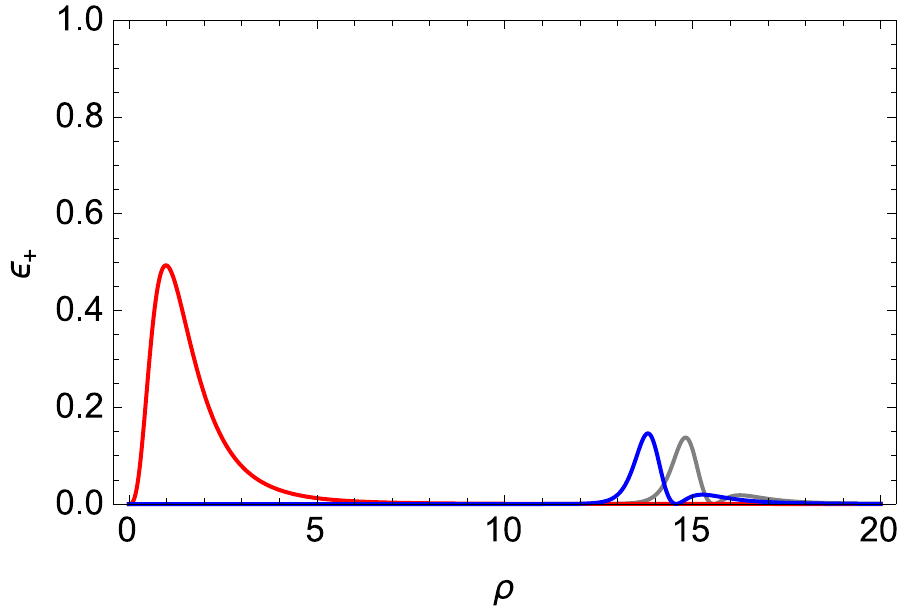}
 \end{minipage} &\ \ \ \ \ \ 
 
  \begin{minipage}[t]{0.3\hsize}
 \centering
\includegraphics[width=5.7cm]{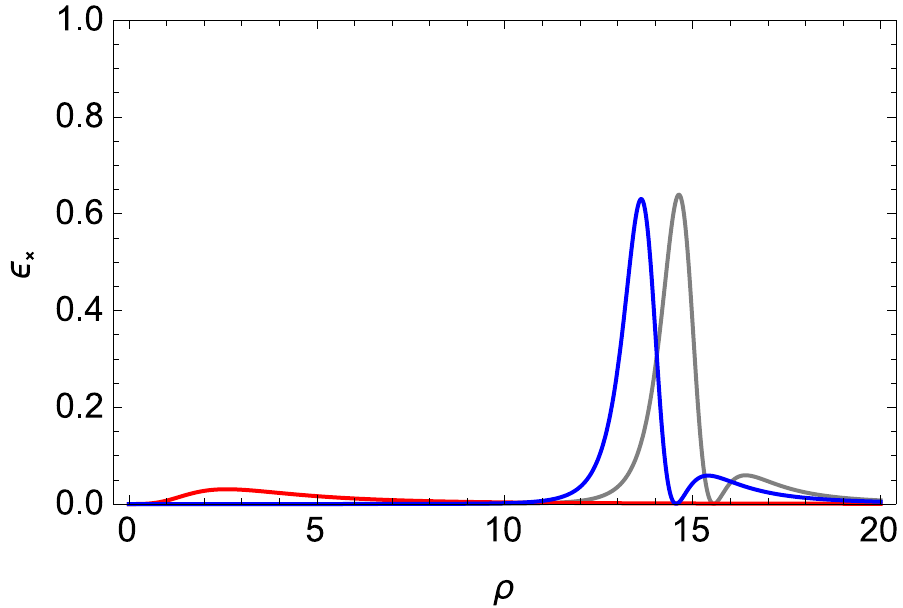}
 \end{minipage}

 \end{tabular}
\caption{Time dependence of the energy densities ${\cal E}$, ${\cal E}_+$,  and ${\cal E}_\times$ for  $(a,c,A,\delta)=(1,2,1,0)$. Each figure displays the snapshots of $\cal E$, $\cal E_+$, and $\cal E_\times$  in the order from left to right, where the gray, red, and blue graphs show the pulses corresponding to $t=-15,0,14$, respectively.
}
\label{fig:Reflection3}
\end{figure}

\begin{figure}[h]
  \begin{tabular}{ccc}

 \begin{minipage}[t]{0.3\hsize}
 \centering
\includegraphics[width=5.7cm]{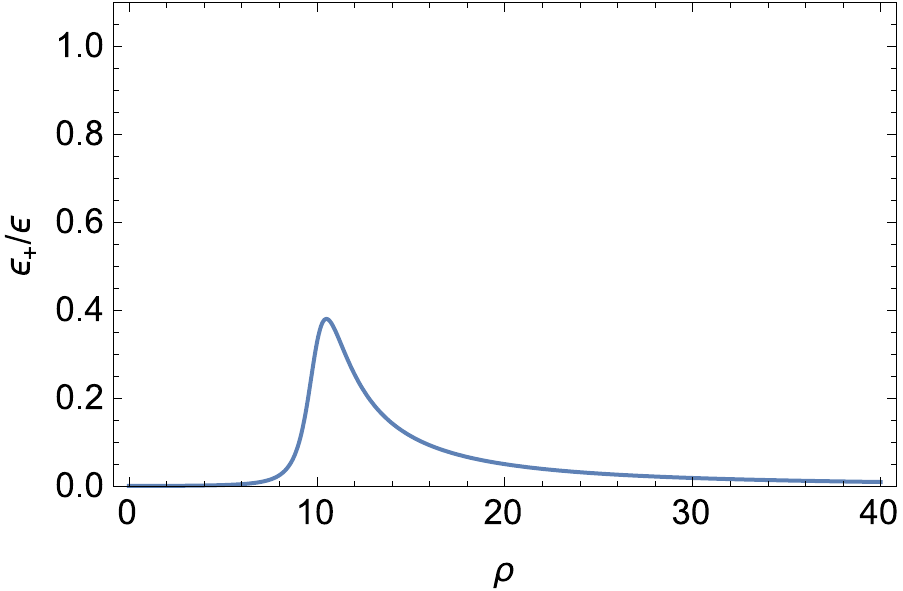}
 \end{minipage} &\ \ \ \ \ \ 
 
  \begin{minipage}[t]{0.3\hsize}
 \centering
\includegraphics[width=5.7cm]{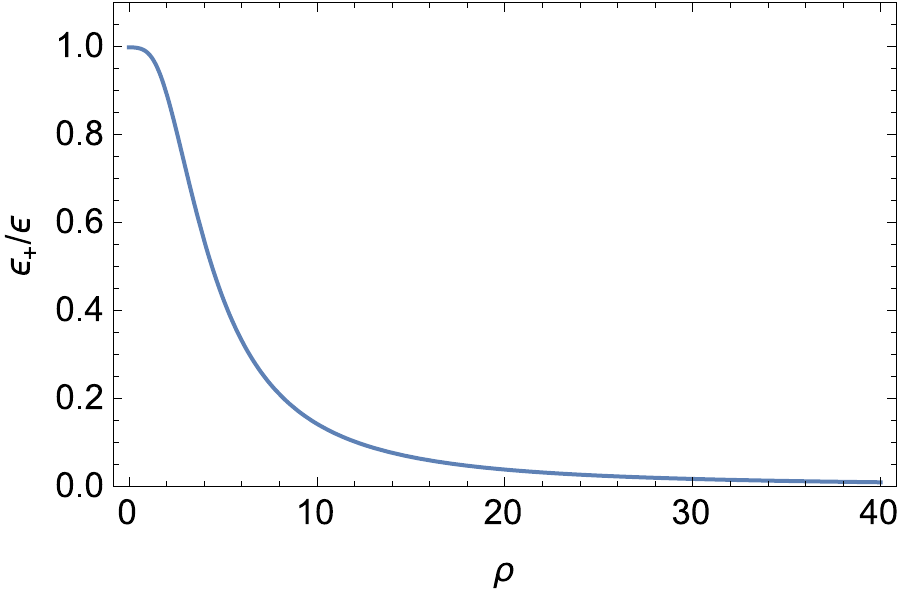}
 \end{minipage} &\ \ \ \ \ \

 \begin{minipage}[t]{0.3\hsize}
 \centering
\includegraphics[width=5.7cm]{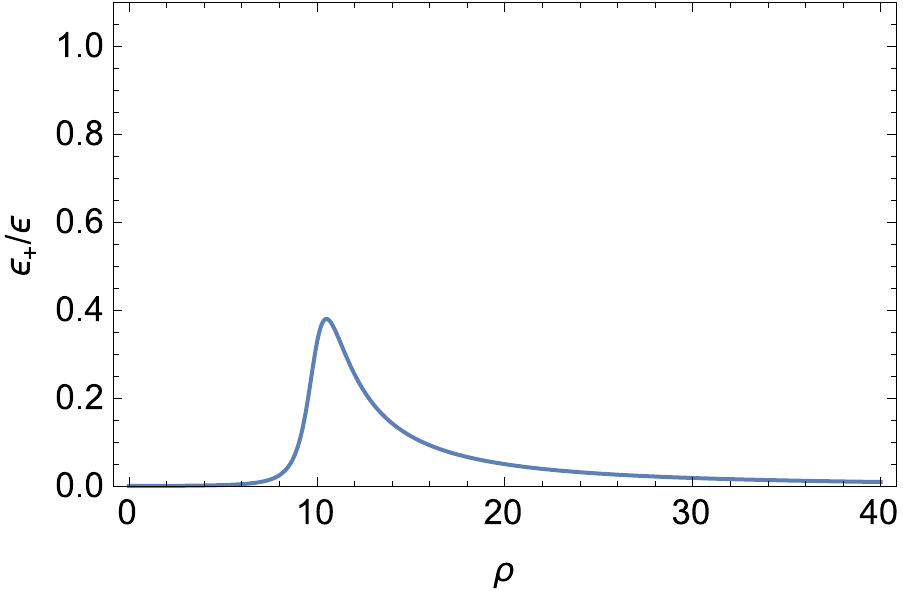}
 \end{minipage}

 \end{tabular}
 \caption{Time dependence of the ratio $\cal E_+/\cal E$ for  $(a,c,A,\delta)=(1,2,1,0)$. In the figures from left to right, the graphs display the ratio at $t=-10,0,10$.}
 \label{fig:Reflection-ratio3}
\end{figure}

\medskip
In contrast, for $A=1$,  displayed numerically in Fig.~\ref{fig:Reflection3}, Fig.~\ref{fig:Reflection-ratio3} and in the right graph in Fig.~\ref{fig:integ1},  the behaviors of gravitational pulses are quite different.  From these figures, it can be seen that the pulse of the $\times$ mode becomes dominant at infinity $\rho=\infty$, whereas at reflection at the axis, the pulse of the $+$ mode only is just temporally generated near the axis but decays far off the axis, and then at last completely vanishes at infinity.  

\medskip
As has been studied by several researchers for two decays~\cite{Piran,tomimatsu},  this conversion between the different polarization modes is considered to be due to the nonlinear term.  
In this present case,  the parameter $A$ physically represents the extent of nonlinear interaction.  In particular, for $A=0$, the solution is reduced to the Einstein-Rosen type, which describes the linear cylindrical waves of only almost  the $+$ mode.  For $A\ll 1$, only the $+$ mode is present at infinity far from the axis of symmetry, whereas it is transformed into the $\times$ mode during the reflection at the axis because the nonlinear self-interaction is rather enhanced.   For $A\simeq 1$, most only the $\times$ mode is present at infinity, whereas  it is transformed into the $+$ mode in the vicinity of the axis.  We have checked that for $\delta\not=0$,  the nonlinear effect, which causes the conversion between the different two modes, becomes much smaller than for $\delta=0$.

\medskip
For simplicity, we put $\delta=0$.  
$\tau$ vanishes at both past and future null infinities, so that it turns out from Eq.~(\ref{eq:condition}) that ${\cal E}_\times>{\cal E}_+$ at null infinities if and only if 
\begin{eqnarray}
\sqrt{2}-1<A<\sqrt{2}+1.
\end{eqnarray}
On the other hand, near the axis, the behavior of $\tau$ is much more complicated, since it has time dependence as
\begin{eqnarray}
\tau \simeq \frac{ac}{t^2+a^2}.
\end{eqnarray}
For example, if the parameters $(a,c,A)$ satisfy
\begin{eqnarray}
\sqrt{2}-1<A<(\sqrt{2}+1)e^{-\frac{2c}{a}}, 
\end{eqnarray}
the $\times$ mode always dominates over the $+$ mode at the axis. 

Also as another example, for the parameter set of $(a,c,A)$ such that
\begin{eqnarray}
(\sqrt{2}+1)e^{-\frac{2c}{a}}<A<\sqrt{2}-1,
\end{eqnarray}
at early time and late time, the $+$ mode is dominant at the axis,  
whereas during the reflection of the pulse, 
the dominant mode 
is, first, the $\times$ mode, then   
varies from the $\times$ mode to the $+$ mode, and returns to the $\times$ mode again.


\subsection{Collision}

First, with the generalized Halilsoy solution obtained in Sec.~\ref{sec:sol1}, let us construct the solutions that represent the collision of multiple gravitational pulses and see numerically how the conversion between both the modes occurs by the nonlinear interaction. 
Note that from the linearity of $\tau$ and the formula~(\ref{eq:omega2}) for $\omega$, it can be easily shown that if $(\tau_1,\omega_1,\gamma_1)$ and $(\tau_2,\omega_2,\gamma_2)$ are arbitrary solutions generated by the harmonic mapping for the same $A$, 
$(\tau_1+\tau_2,\omega_1+\omega_2,\gamma)$ is also a solution, where $\gamma$ is determined by replacing $(\tau,\omega)$ with $(\tau_1+\tau_2,\omega_1+\omega_2)$ in Eq.~(\ref{eq:gamma1}).
For instance, provided that $(\tau(t),\omega(t),\gamma(t))$ is a solution with the parameters $(a,c,A,\delta)$ generated by the harmonic mapping, one can superpose two different solutions obtained by time shifts $t_1$, $t_2$: $(\tau(t-t_1),\omega(t-t_1),\gamma(t-t_1))$ and $(\tau(t-t_2),\omega(t-t_2),\gamma(t-t_2))$ for the same $(a,c,A,\delta)$. 
Furthermore, in general, one can superpose an arbitrary number of different solutions with arbitrary parameters $(a_i,c_i,\delta_i)$ and time shifts $t_i$: 
$(\tau(t-t_i),\omega(t-t_i),\gamma(t-t_i),a_i,c_i,\delta_i)$ for the same $A$, 
so that by arranging the time shifts appropriately, the collisional solutions required are obtained.

\begin{figure}[h]
  \begin{tabular}{ccc}
 \begin{minipage}[t]{0.3\hsize}
 \centering
\includegraphics[width=5.7cm]{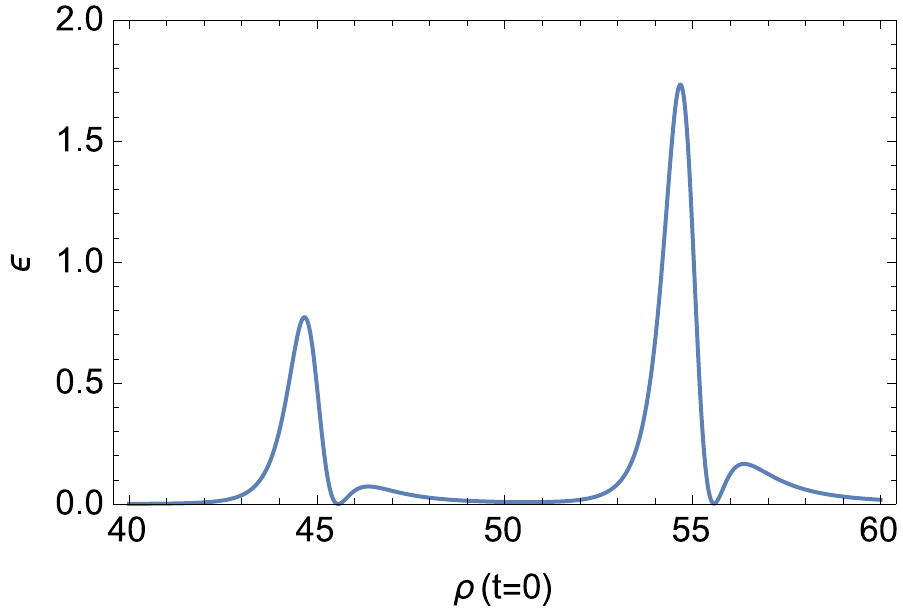}
 \end{minipage} &\ \ \ \ \ \ 

 \begin{minipage}[t]{0.3\hsize}
 \centering
\includegraphics[width=5.7cm]{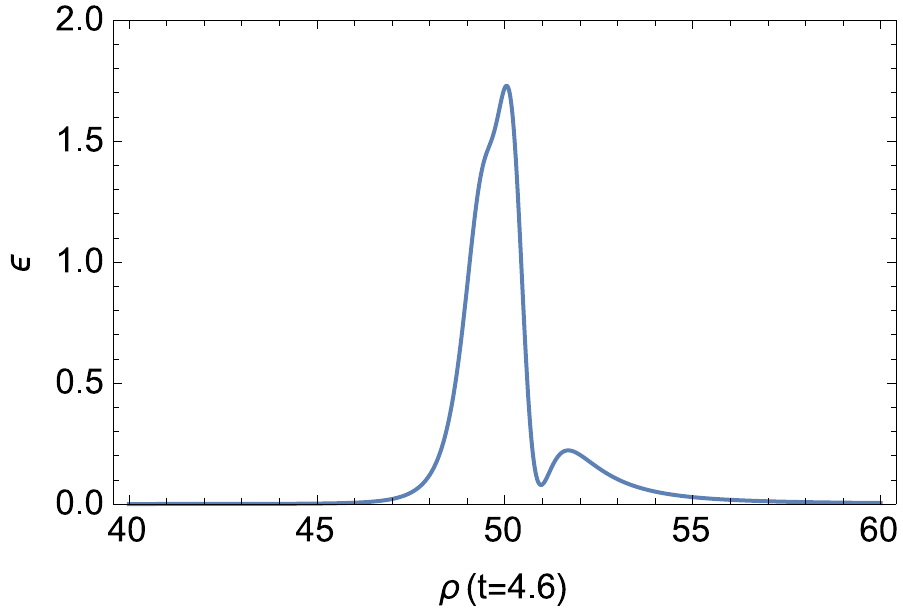}
 \end{minipage} &\ \ \ \ \ \ 
 
  \begin{minipage}[t]{0.3\hsize}
 \centering
\includegraphics[width=5.7cm]{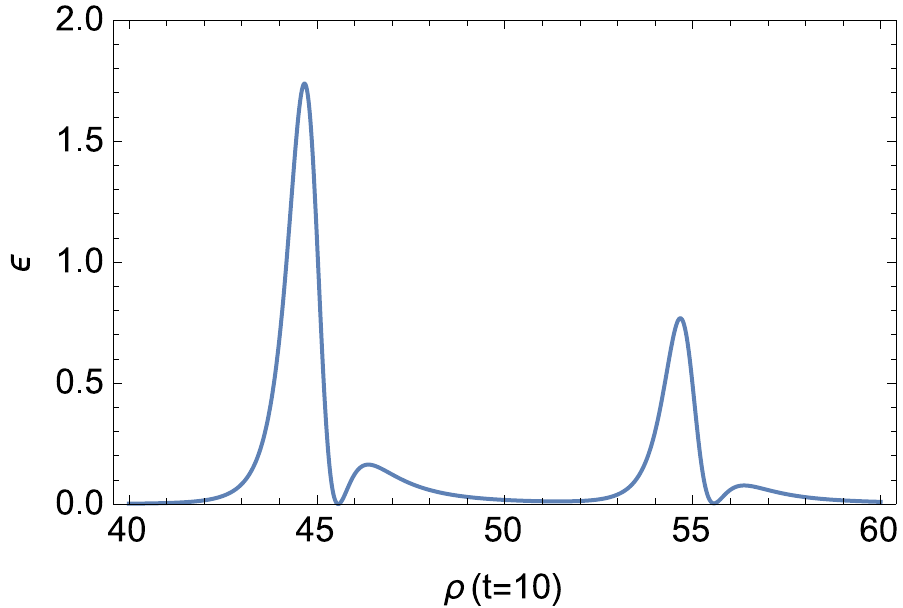}

 \end{minipage}\\ &
 \\

 \begin{minipage}[t]{0.3\hsize}
 \centering
\includegraphics[width=5.7 cm]{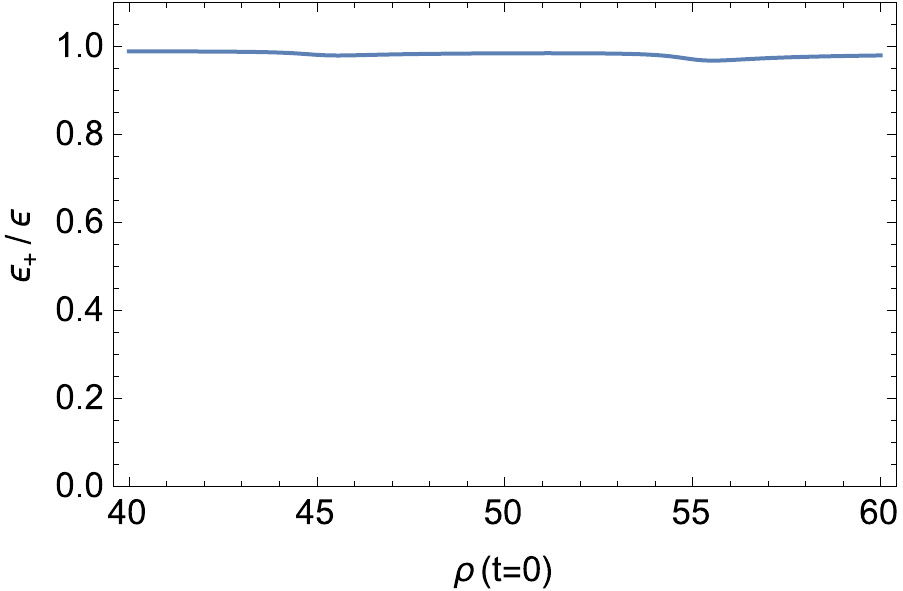}
 \end{minipage}&\ \ \ \ \ \ 
 
 \begin{minipage}[t]{0.3\hsize}
 \centering
\includegraphics[width=5.7cm]{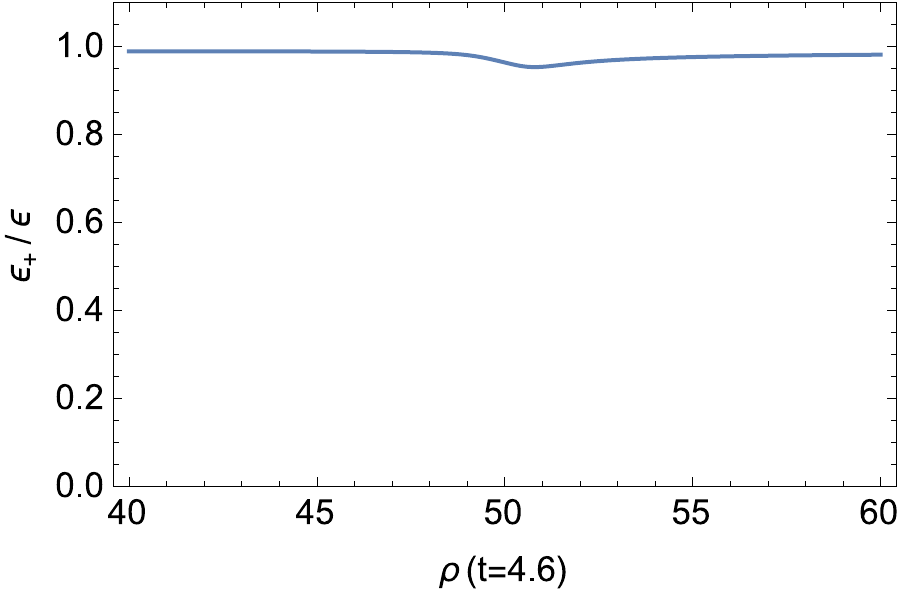}
 \end{minipage} &\ \ \ \ \ \ 
 
  \begin{minipage}[t]{0.3\hsize}
 \centering
\includegraphics[width=5.7cm]{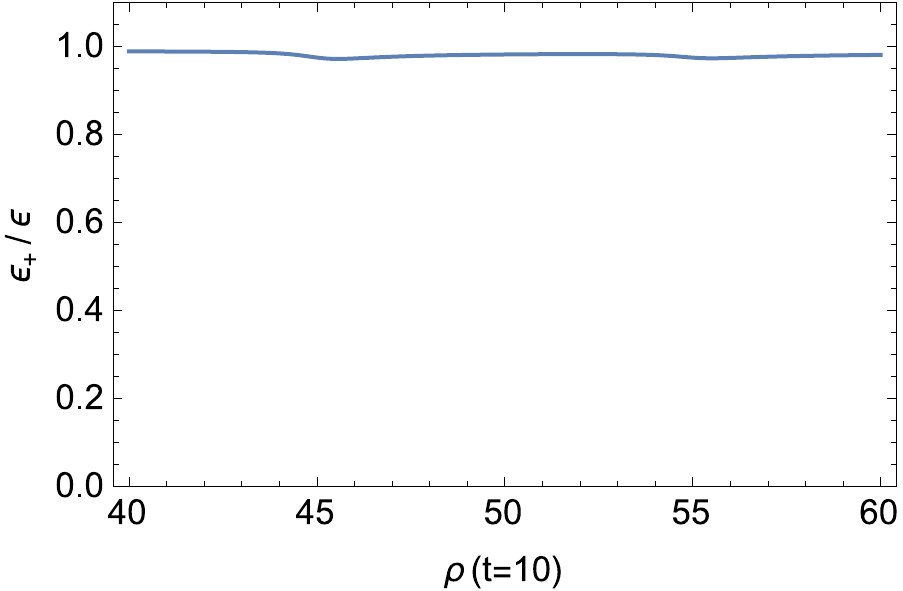}
 \end{minipage}

 \end{tabular}
 \caption{The upper graphs show the snapshots of pulses at the collision for $(a_1,a_2,c_1,c_2,t_1,t_2,A,\delta)=(1,1,2,3,-45,55,0.05,0)$, where  from left to right, the three graphs correspond to $t=0,4.6,10$.   The lower graphs show the ratio $\cal E_+/\cal E$ at each time.}
 \label{fig:collision1}
\end{figure}

\begin{figure}[h]

  \begin{tabular}{ccc}
 \begin{minipage}[t]{0.45\hsize}
 \centering
\includegraphics[width=8cm]{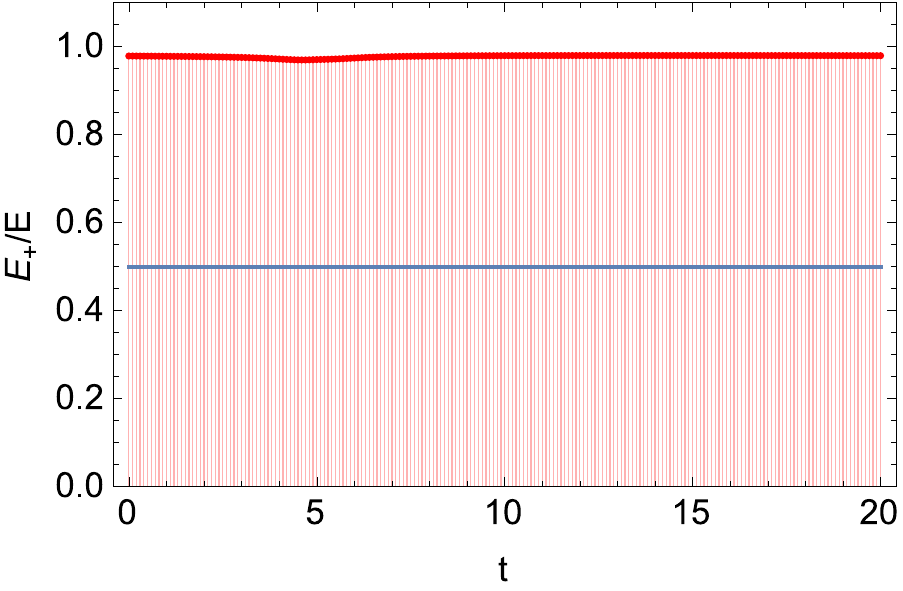}
 \end{minipage} &\ \ \ \ \ \ 
 
 \begin{minipage}[t]{0.45\hsize}
 \centering
\includegraphics[width=8cm]{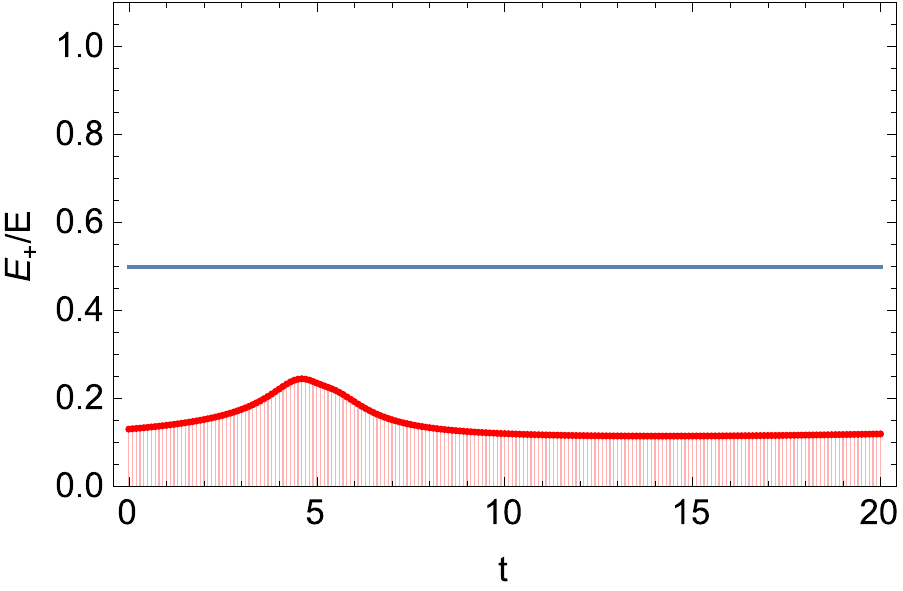}
 \end{minipage} &\ \ \ \ \ \

 \end{tabular}
\caption{ The time dependence of the ratio $E_+(t,1000)/E(t,1000)$
 when an outgoing pulse with $(a_1,c_1,t_1)=(1,2,-45)$ and an ingoing pulse with $(a_2,c_2,t_2)=(1,3,55)$ collide at $t\simeq 5$, where the left and right graphs display the ratio for $(A,\delta)=(0.05,0)$ and   $(A,\delta)=(1,0)$, respectively.   }
\label{fig:integcollision1-1}
\end{figure}

In the following, we set $\delta$ to 0 for simplicity. 
The upper graphs in Fig.\ref{fig:collision1} show the collision of an outgoing small pulse  with $(t_1,a_1,c_1)=(-45,1,2)$ and an ingoing large pulse with $(t_2,a_2,c_2)=(55,1,3)$ for $A=0.05$.
The three graphs are taken at $t=0,4.6,10$, respectively. 
The lower graphs show how the ratio of the energy density of the $+$ mode to the total energy density, $\cal E_+/\cal E$, changes in the collision process. 
It is numerically shown that before the collision, the ratio $\cal E_+/\cal E$ is almost $1$, whereas only at the moment they collide, it is slightly decreased, and after the collision,  it again returns to $1$. Moreover,  the left graph in Fig.\ref{fig:integcollision1-1} shows the time development of 
$E_+(t,1000)/E(t,1000)$.
 It can be seen that only when two pulses with the $+$ mode collide at $t\simeq 4.6$, just a little amount of energy of the $+$ mode is transformed into the energy of the $\times$ mode.

\medskip

\begin{figure}[h]
  \begin{tabular}{ccc}
 \begin{minipage}[t]{0.3\hsize}
 \centering
\includegraphics[width=5.7cm]{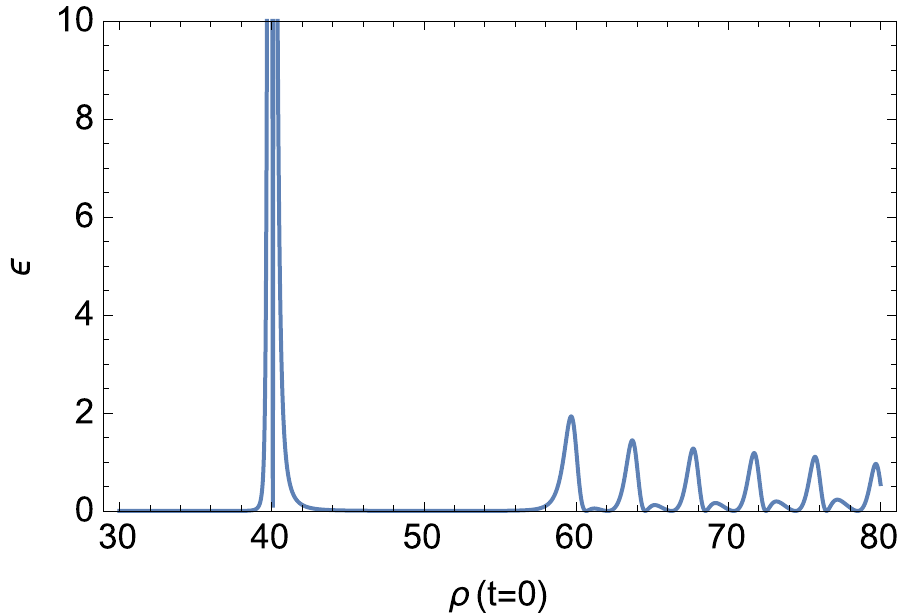}
 \end{minipage} &\ \ \ \ \ \ 
 
 \begin{minipage}[t]{0.3\hsize}
 \centering
\includegraphics[width=5.7cm]{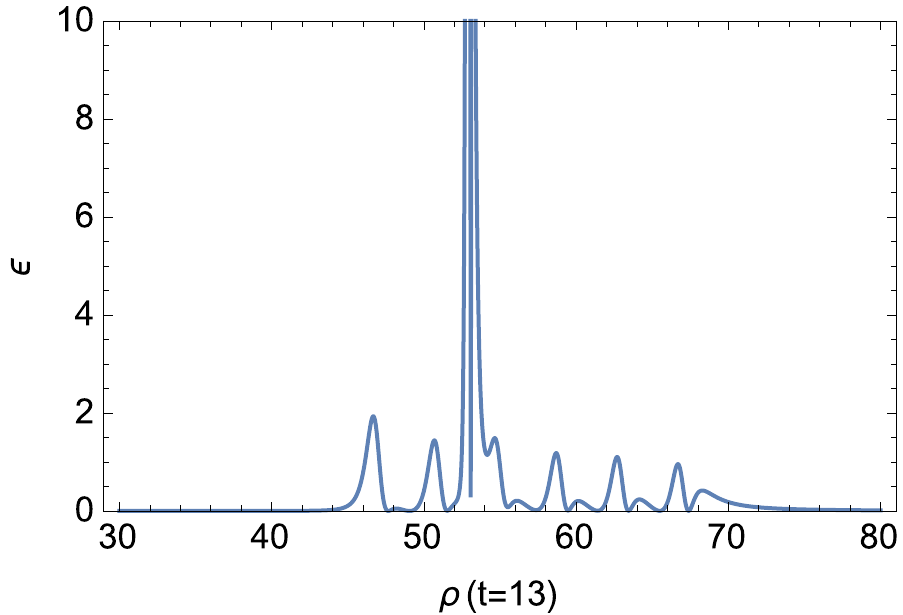}
 \end{minipage} &\ \ \ \ \ \ 
 
  \begin{minipage}[t]{0.3\hsize}
 \centering
\includegraphics[width=5.7cm]{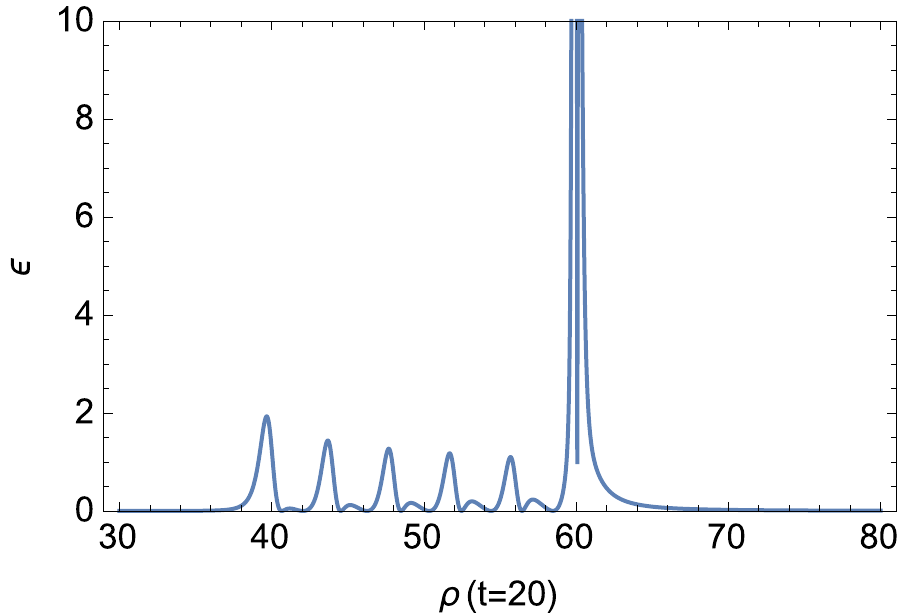}
 \end{minipage}\\ &
 \\

 \begin{minipage}[t]{0.3\hsize}
 \centering
\includegraphics[width=5.7cm]{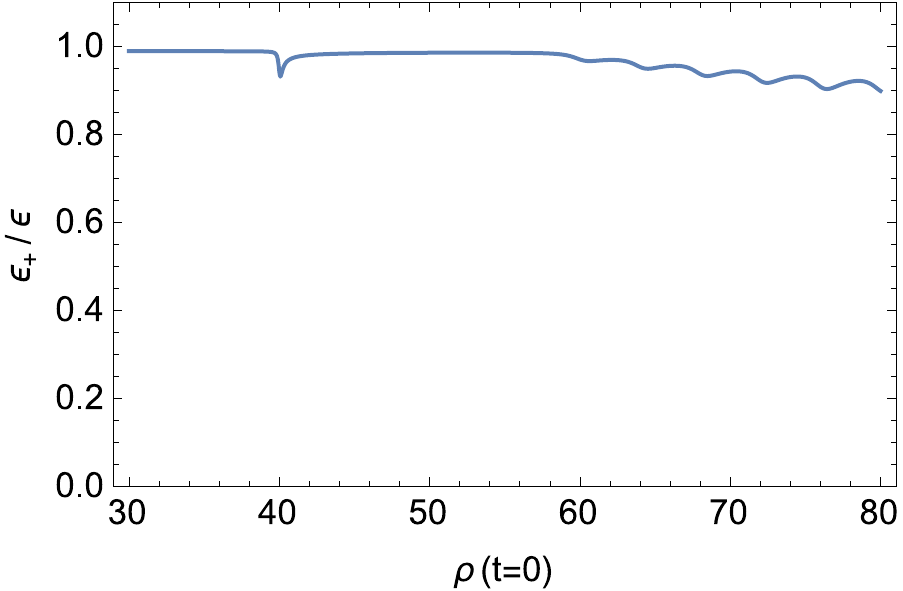}
 \end{minipage}&\ \ \ \ \ \ 
 
 \begin{minipage}[t]{0.3\hsize}
 \centering
\includegraphics[width=5.7cm]{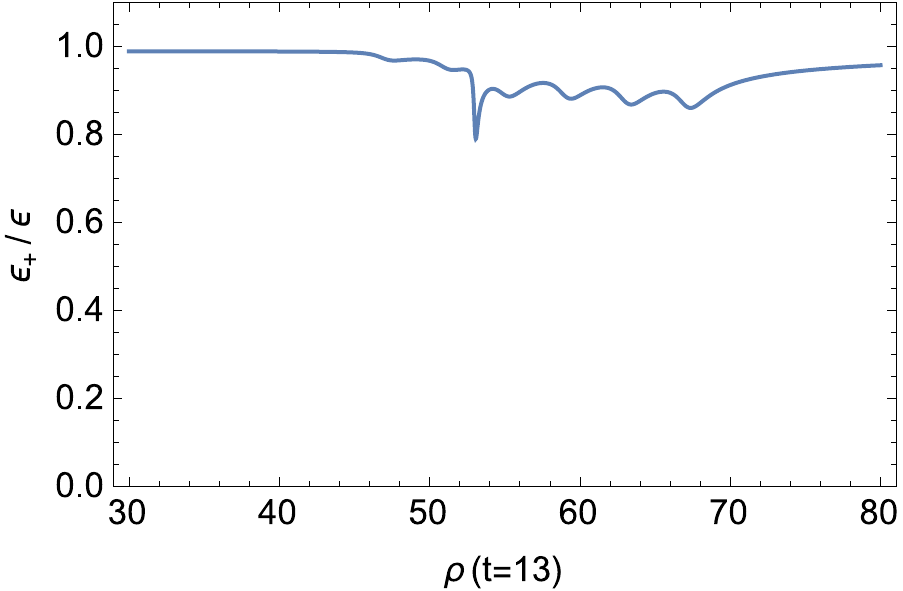}
 \end{minipage} &\ \ \ \ \ \ 
 
  \begin{minipage}[t]{0.3\hsize}
 \centering
\includegraphics[width=5.7cm]{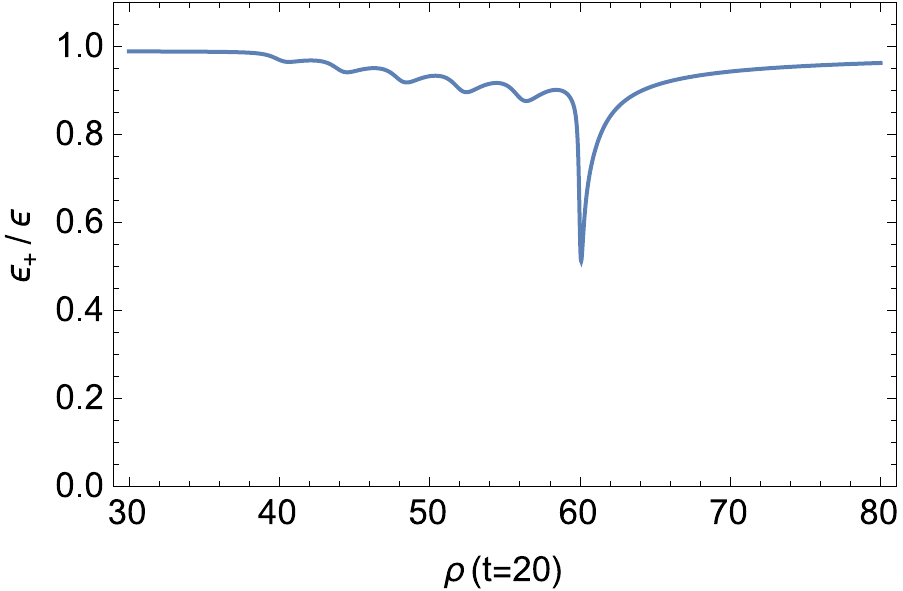}
 \end{minipage}

 \end{tabular}
\caption{The snapshots at the collision of a single large gravitational pulse with $(t_0,a_0,c_0)=(-40,1/7,2)$ and six small gravitational pulses with $(t_i,a_i,c_i)=(56+4i,1,3)\ (i=1,\cdots,6)$ for $A=0.05$ and $\delta=0$, where  from left to right, the three graphs correspond to $t=0,13,20$.   The lower graphs show the ratio $\cal E_+/\cal E$ at each time.}
 \label{fig:collision6a}
\end{figure}

\begin{figure}[h]
  \begin{tabular}{ccc}
 \begin{minipage}[t]{0.3\hsize}
 \centering
\includegraphics[width=5.7cm]{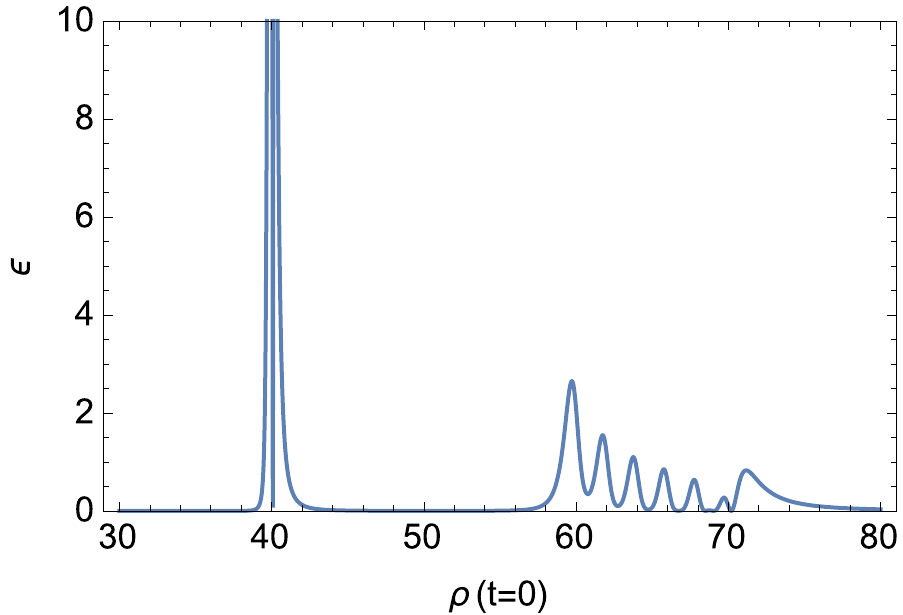}
 \end{minipage} &\ \ \ \ \ \ 
 
 \begin{minipage}[t]{0.3\hsize}
 \centering
\includegraphics[width=5.7cm]{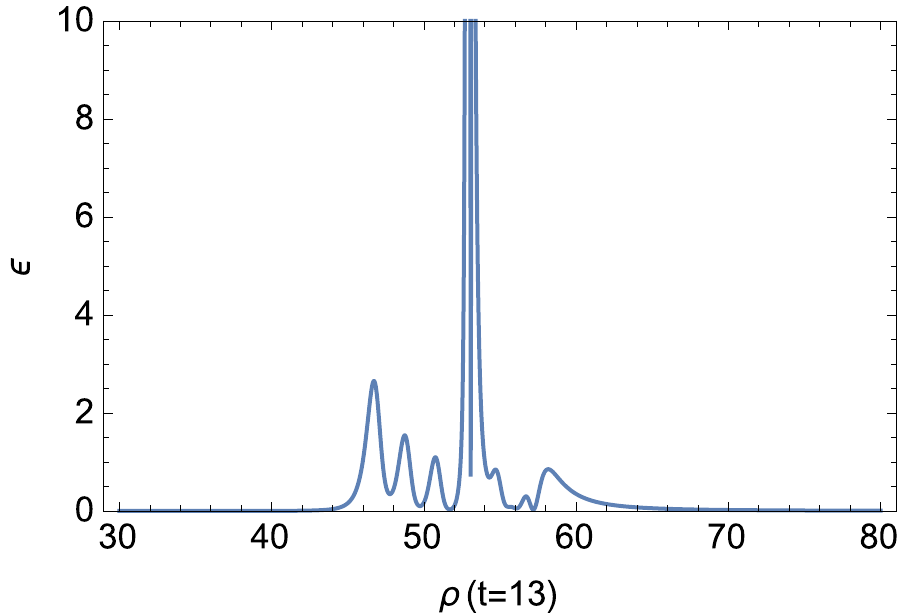}
 \end{minipage} &\ \ \ \ \ \ 
 
  \begin{minipage}[t]{0.3\hsize}
 \centering
\includegraphics[width=5.7cm]{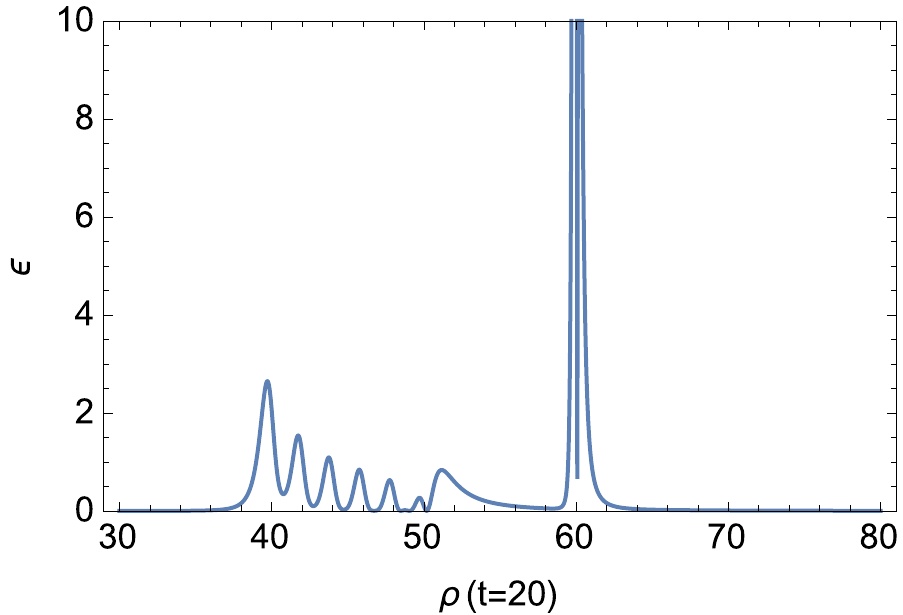}
 \end{minipage}\\ &
 \\

 \begin{minipage}[t]{0.3\hsize}
 \centering
\includegraphics[width=5.7cm]{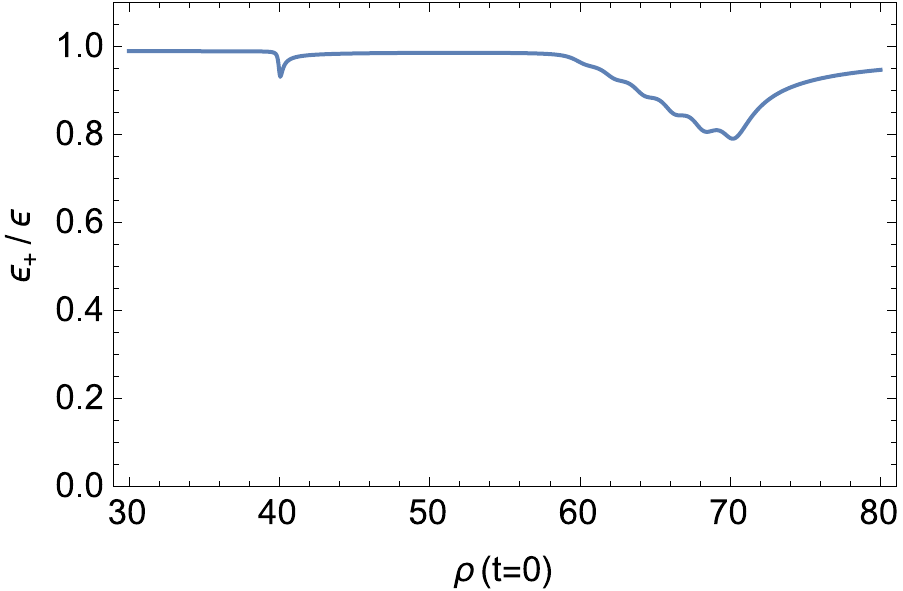}
 \end{minipage}&\ \ \ \ \ \ 
 
 \begin{minipage}[t]{0.3\hsize}
 \centering
\includegraphics[width=5.7cm]{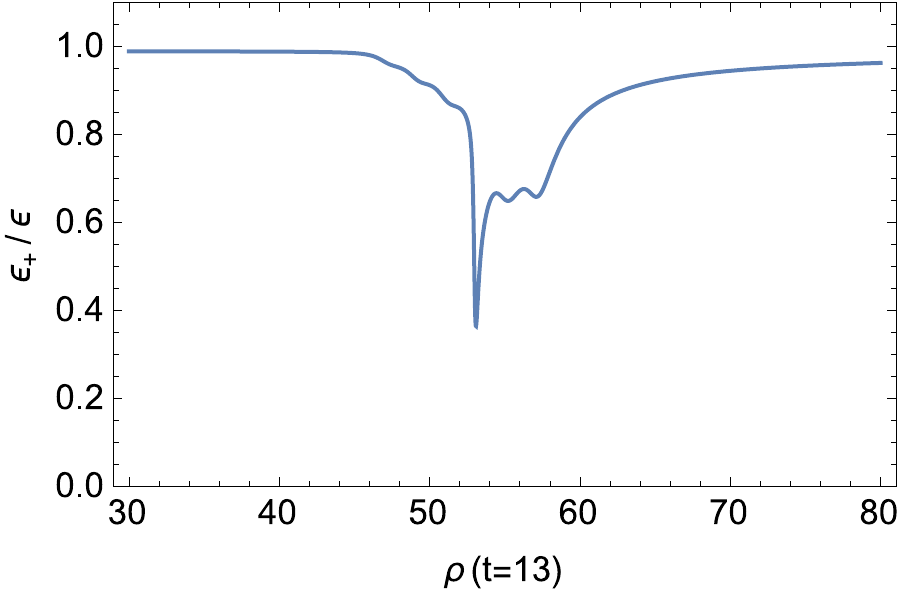}
 \end{minipage} &\ \ \ \ \ \ 
 
  \begin{minipage}[t]{0.3\hsize}
 \centering
\includegraphics[width=5.7cm]{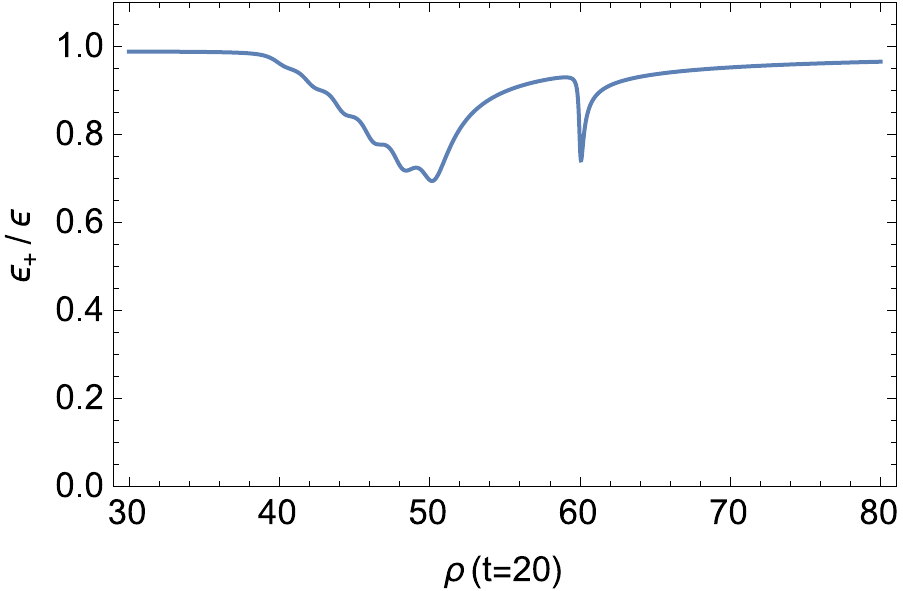}
 \end{minipage}

 \end{tabular}
\caption{The upper graphs show the snapshots at the collision of a single large gravitational pulse with $(t_0,a_0,c_0)=(-40,1/7,2)$ and six small gravitational pulses with $(t_i,a_i,c_i)=(58+2i,1,3)\ (i=1,\cdots,6)$ for $A=0.05$ and $\delta=0$, where  from left to right, the three graphs correspond to $t=0,13,20$.   The lower graphs show the ratio $\cal E_+/\cal E$ at each time.}
 \label{fig:collision6b}
\end{figure}

\begin{figure}[h]

  \begin{tabular}{ccc}
 \begin{minipage}[t]{0.45\hsize}
 \centering
\includegraphics[width=8cm]{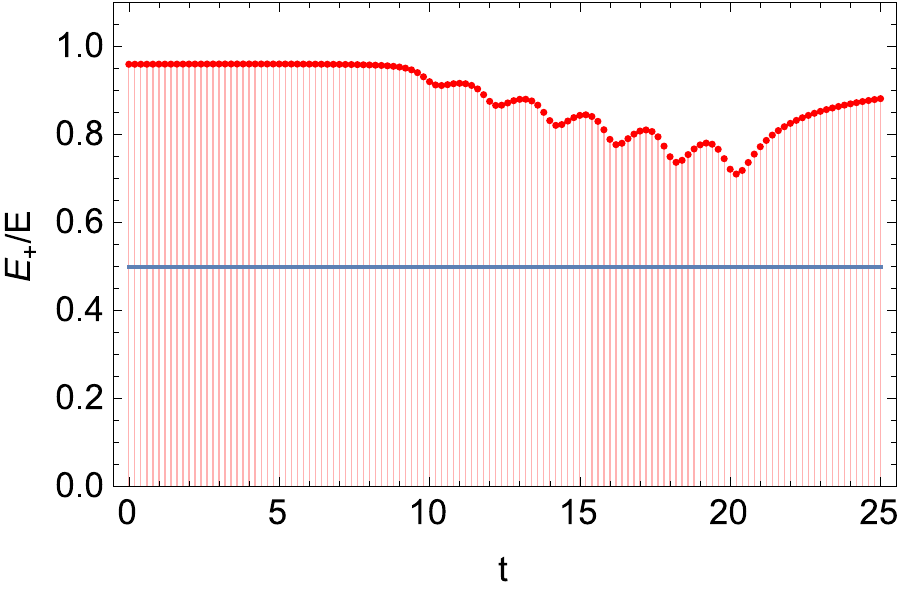}
 \end{minipage} &\ \ \ \ \ \ 
 
 \begin{minipage}[t]{0.45\hsize}
 \centering
\includegraphics[width=8cm]{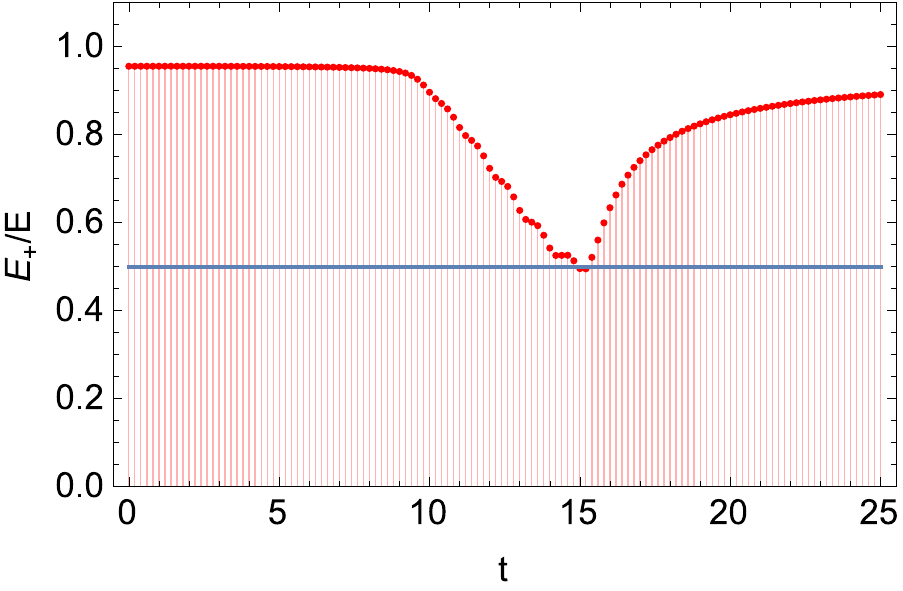}
 \end{minipage} &\ \ \ \ \ \

 \end{tabular}
\caption{The time dependence of the ratio $E_+(t,1000)/E(t,1000)$
 when an outgoing pulse with $(a_0,c_0,t_0)=(1/7,2,-40)$ and six ingoing pulses with $(a_i,c_i)=(1,3)$  collide for $(A=0.05,0)$, where the left and right graphs display the ratio for $t_i=56+4i$ and   $t_i=58+2i$ $(i=1,\ldots,6)$, respectively.  
}
\label{fig:integ1-6a}
\end{figure}

Next, we consider the collision of a single outgoing large gravitational pulse with $(t_0,a_0,c_0)$ and six ingoing small gravitational pulses with $(t_i,a_i,c_i)\ (i=1,\cdots,6)$  and see how they mutually interact. 
The upper graphs in Fig.\ref{fig:collision6a} (Fig.\ref{fig:collision6b}) display the snapshots  of the total energy density ${\cal E}$ at each time of $t=0,13,20$ when a large pulse with $(t_0,a_0,c_0)=(-40,1/7,2)$ and small pulses with $(t_i,a_i,c_i)=(56+4i,1,3)$ [$(t_i,a_i,c_i)=(58+2i,1,3)$ in Fig.\ref{fig:collision6b}] for $A=0.05$, and the lower graphs show how the ratio ${\cal E}_+/{\cal E}$ changes in the collision process for the corresponding parameters. 
It is shown that when they collide, the $\times$ mode is temporarily increased but  is decreased shortly afterward, and the $+$ mode becomes dominant at late time.   Also the left and right graphs in Fig.\ref{fig:integ1-6a} display  the time development of $E_+(t,1000)/E(t,1000)$ for $0\le t\le 20$ for $(a_0,a_i,c_0,c_i,A,\delta,t_0,t_i)=(1/7,1,2,3,1,0,-40,56+4i)$ and $(a_0,a_i,c_0,c_i,A,\delta,t_0,t_i)=(1/7,1,2,3,1,0,-40,58+2i)$, respectively. Only at the moment the pulses collide, the energy of the $+$ mode is converted to that of the $\times$ mode, while the energy of the $+$ mode comes to dominate that of the $\times$ mode once they go away from each other.  For $\delta\not=0$,  it can be numerically confirmed that the rate of the conversion from the $+$ mode to the $\times$ mode becomes smaller than for $\delta=0$.

\begin{figure}[h]
  \begin{tabular}{ccc}
 \begin{minipage}[t]{0.3\hsize}
 \centering
\includegraphics[width=5.7cm]{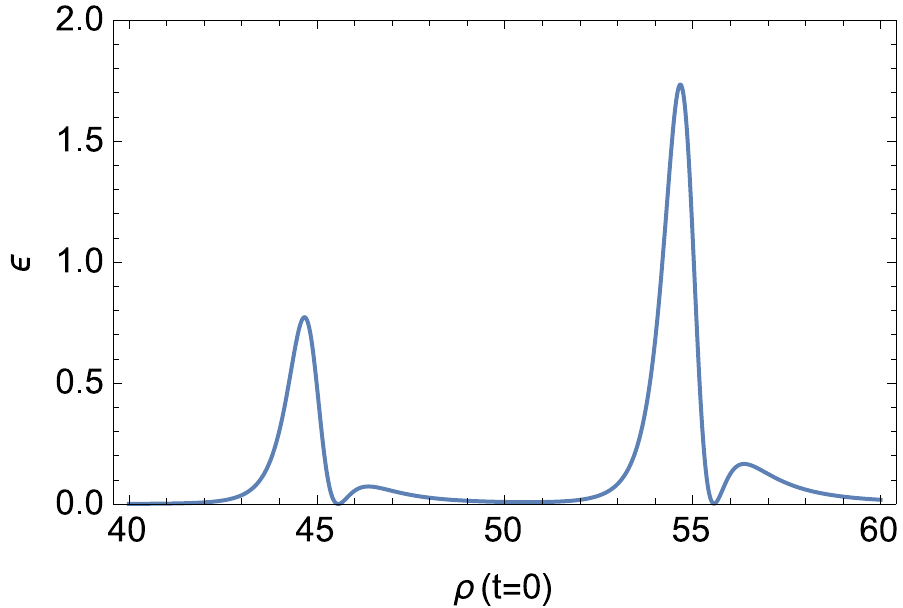}
 \end{minipage} &\ \ \ \ \ \ 
 
 \begin{minipage}[t]{0.3\hsize}
 \centering
\includegraphics[width=5.7cm]{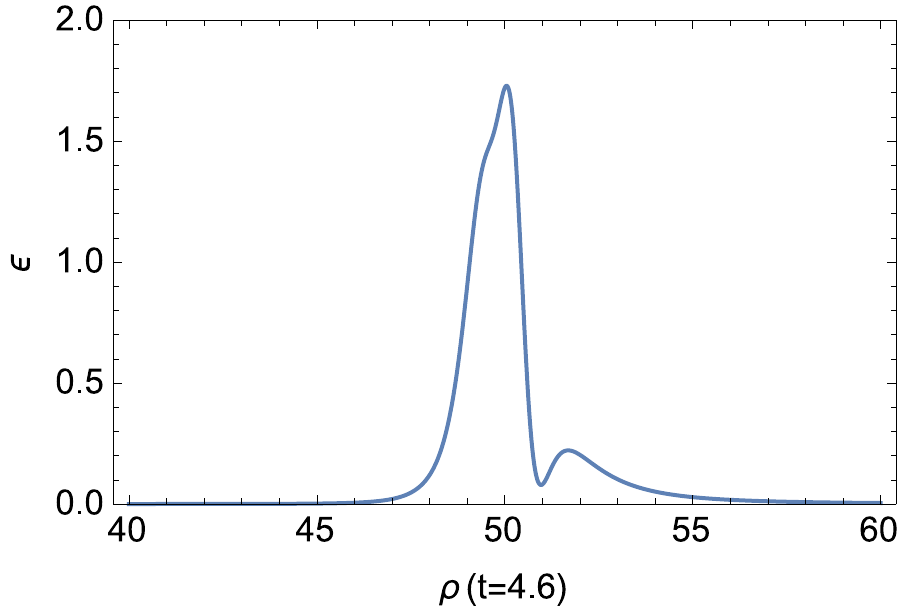}
 \end{minipage} &\ \ \ \ \ \ 
 
  \begin{minipage}[t]{0.3\hsize}
 \centering
\includegraphics[width=5.7cm]{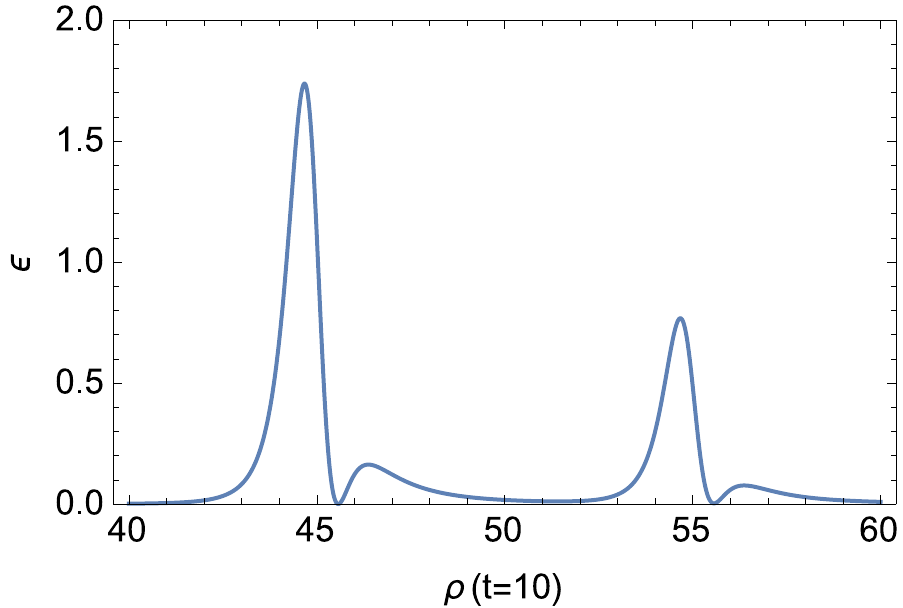}
 \end{minipage}\\ &
 \\

 \begin{minipage}[t]{0.3\hsize}
 \centering
\includegraphics[width=5.7cm]{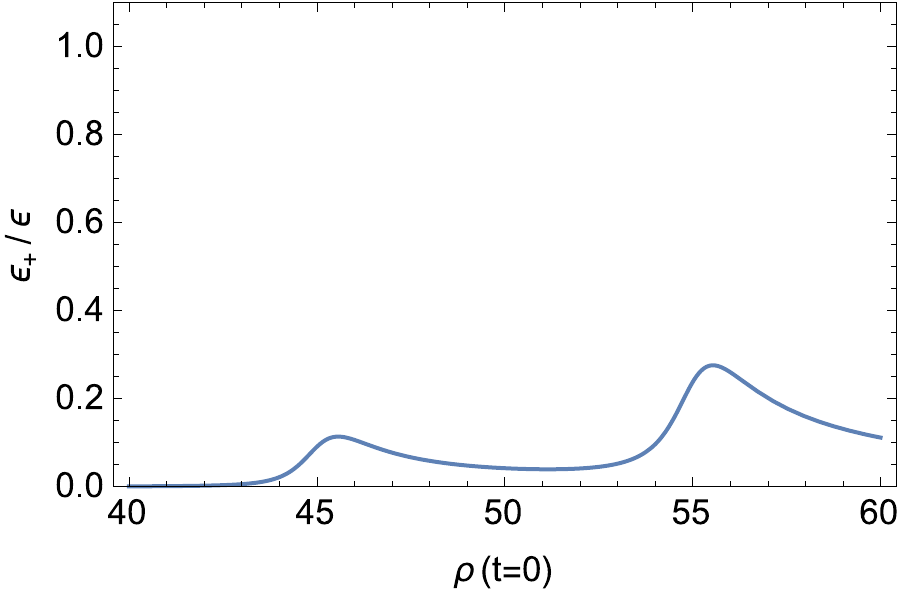}
 \end{minipage}&\ \ \ \ \ \ 
 
 \begin{minipage}[t]{0.3\hsize}
 \centering
\includegraphics[width=5.7cm]{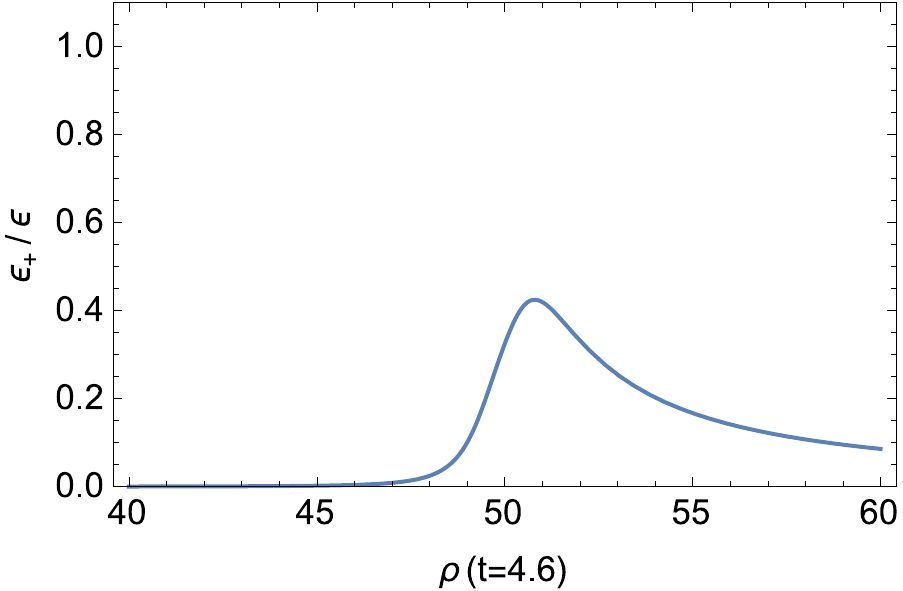}
 \end{minipage} &\ \ \ \ \ \ 
 
  \begin{minipage}[t]{0.3\hsize}
 \centering
\includegraphics[width=5.7cm]{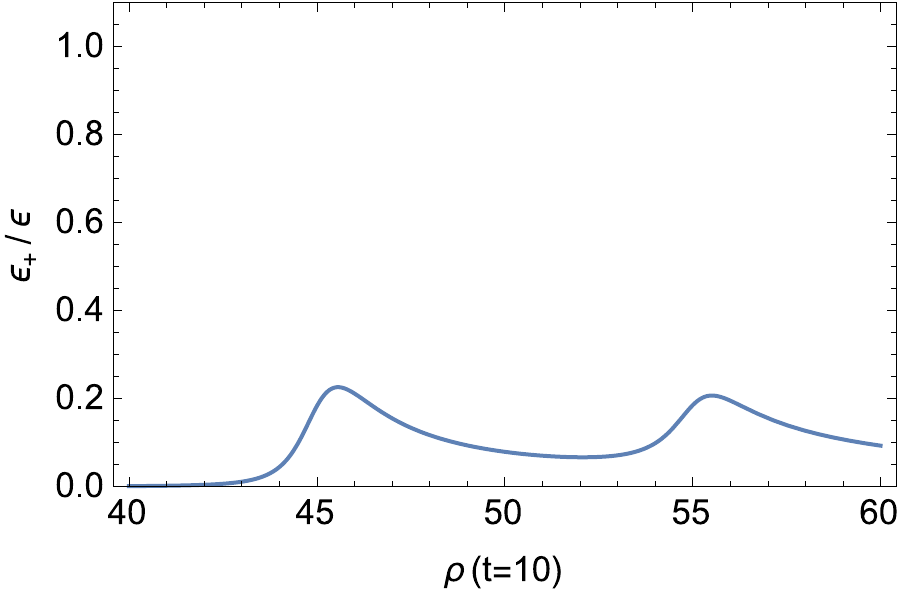}
 \end{minipage}

 \end{tabular}
 \caption{The upper graphs show the snapshots of pulses at the collision for $(a_1,a_2,c_1,c_2,t_1,t_2,A,\delta)=(1,1,2,3,-45,55,1,0)$, where  from left to right, the three graphs correspond to $t=0,4.6,10$.  The lower graphs show the ratio $\cal E_+/\cal E$ at each time.}
 \label{fig:collision2}
\end{figure}

\medskip
For $A\simeq 1$, the situation is considerably different. 
The upper ones in Fig.\ref{fig:collision2} show the snapshots of $t=0,4.6,10$ at the collision of an outgoing small pulse  with $(t_1,a_1,c_1)=(-45,1,2)$ and an ingoing large pulse with $(t_2,a_2,c_2)=(55,1,3)$ for $A=1$, and the lower ones show the ratio $\cal E_+/\cal E$ at the respective times.  From the right graph in Fig.~\ref{fig:integcollision1-1}, it can be observed that the $\times$ mode is dominant far away from the axis, whereas the $\times$ mode is transformed into the $+$ mode only when two pulses meet $t\simeq4.6$.

\begin{figure}[h]
  \begin{tabular}{ccc}
 \begin{minipage}[t]{0.3\hsize}
 \centering
\includegraphics[width=5.7cm]{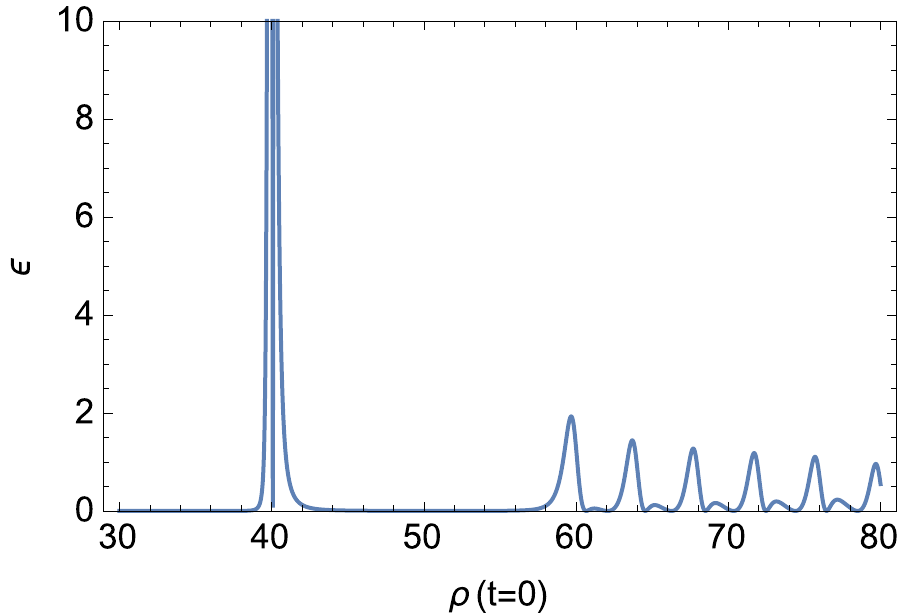}
 \end{minipage} &\ \ \ \ \ \ 
 
 \begin{minipage}[t]{0.3\hsize}
 \centering
\includegraphics[width=5.7cm]{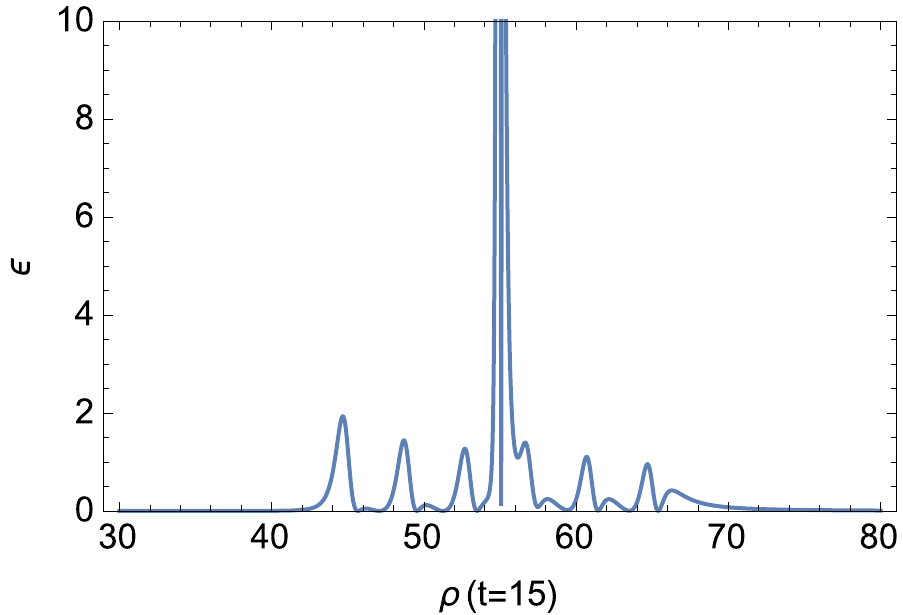}
 \end{minipage} &\ \ \ \ \ \ 
 
  \begin{minipage}[t]{0.3\hsize}
 \centering
\includegraphics[width=5.7cm]{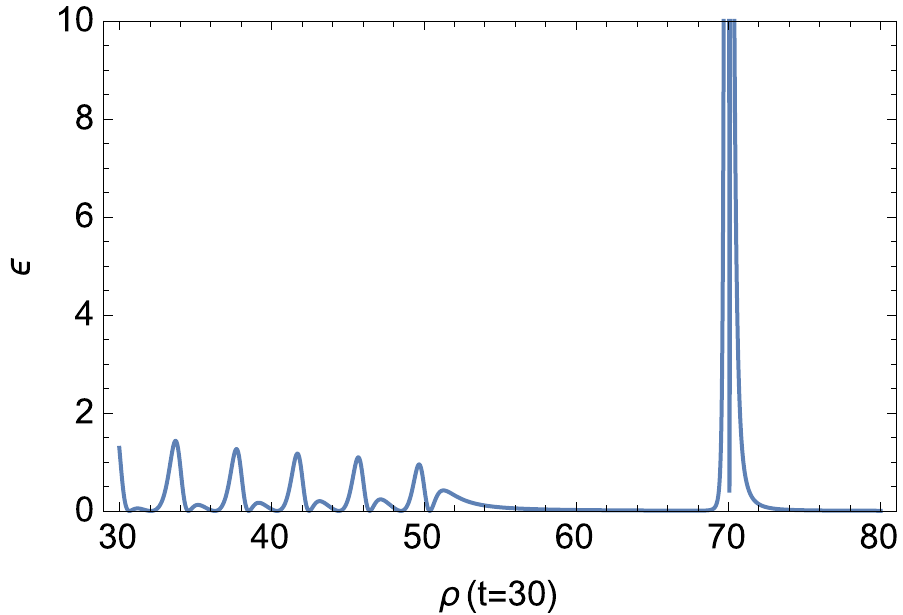}
 \end{minipage}\\ &
 \\

 \begin{minipage}[t]{0.3\hsize}
 \centering
\includegraphics[width=5.7cm]{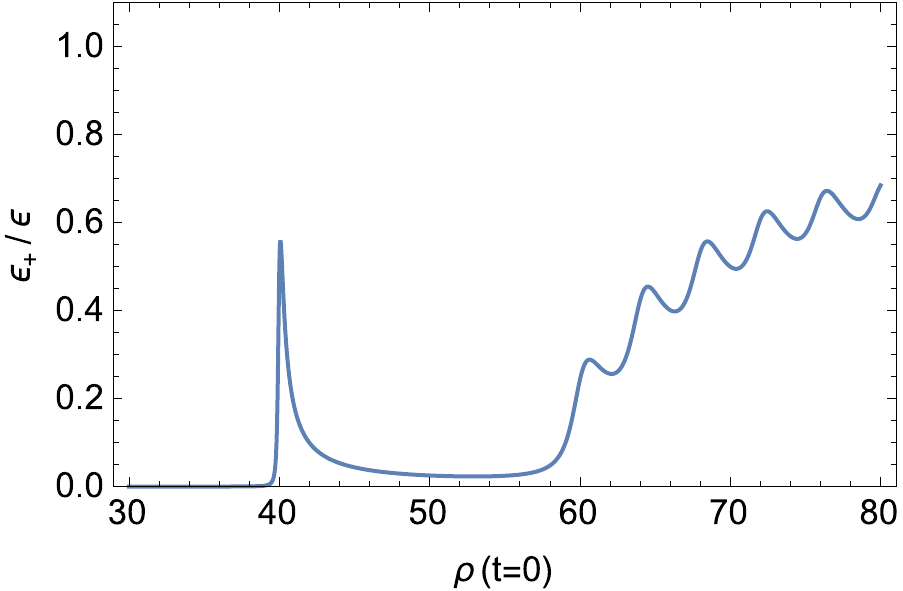}
 \end{minipage}&\ \ \ \ \ \ 
 
 \begin{minipage}[t]{0.3\hsize}
 \centering
\includegraphics[width=5.7cm]{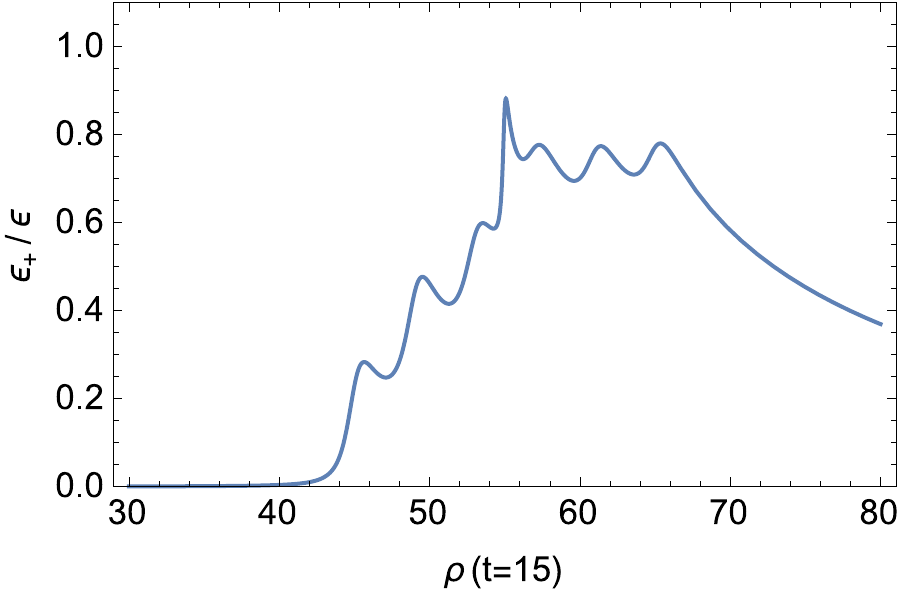}
 \end{minipage} &\ \ \ \ \ \ 
 
  \begin{minipage}[t]{0.3\hsize}
 \centering
\includegraphics[width=5.7cm]{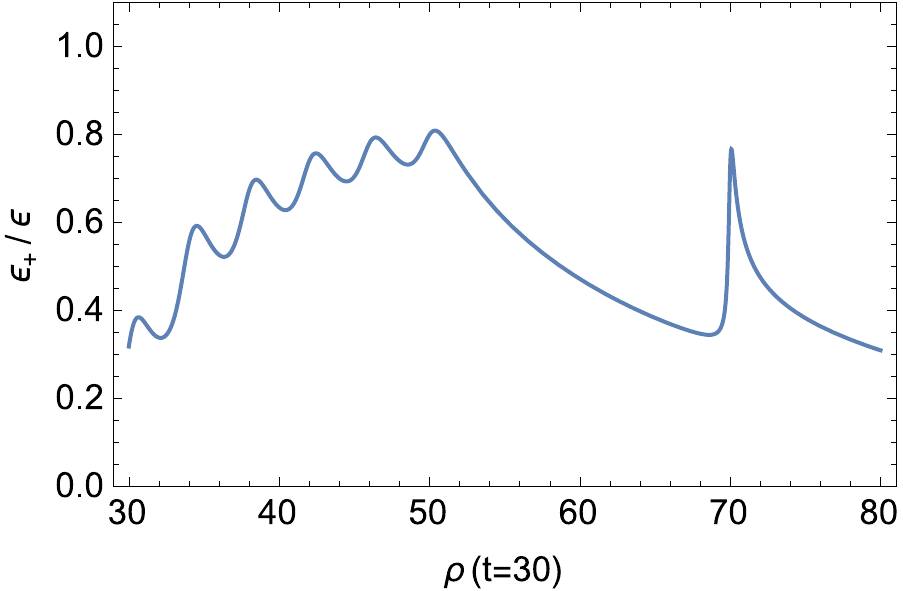}
 \end{minipage}

 \end{tabular}
\caption{The upper graphs show the snapshots at the collision of a single large gravitational pulse with $(t_0,a_0,c_0)=(-40,1/7,2)$ and six small gravitational pulses with $(t_i,a_i,c_i)=(56+4i,1,3)\ (i=1,\cdots,6)$ for $A=1$ and $\delta=0$, where  from left to right, the three graphs correspond to $t=0,15,30$.   The lower graphs show the ratio $\cal E_+/\cal E$ at each time.}
 \label{fig:collision6c}
\end{figure}

\begin{figure}[h]
  \begin{tabular}{ccc}
 \begin{minipage}[t]{0.3\hsize}
 \centering
\includegraphics[width=5.7cm]{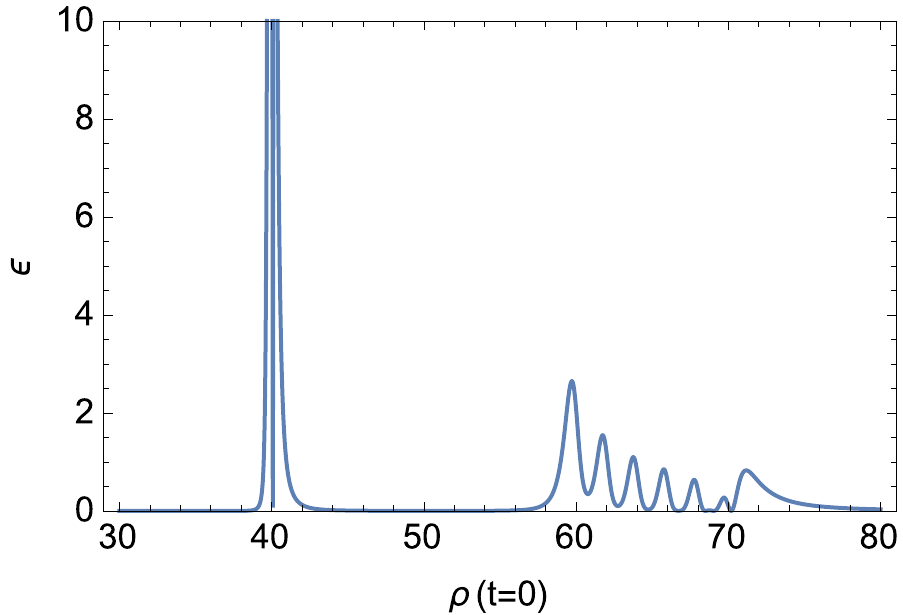}
 \end{minipage} &\ \ \ \ \ \ 
 
 \begin{minipage}[t]{0.3\hsize}
 \centering
\includegraphics[width=5.7cm]{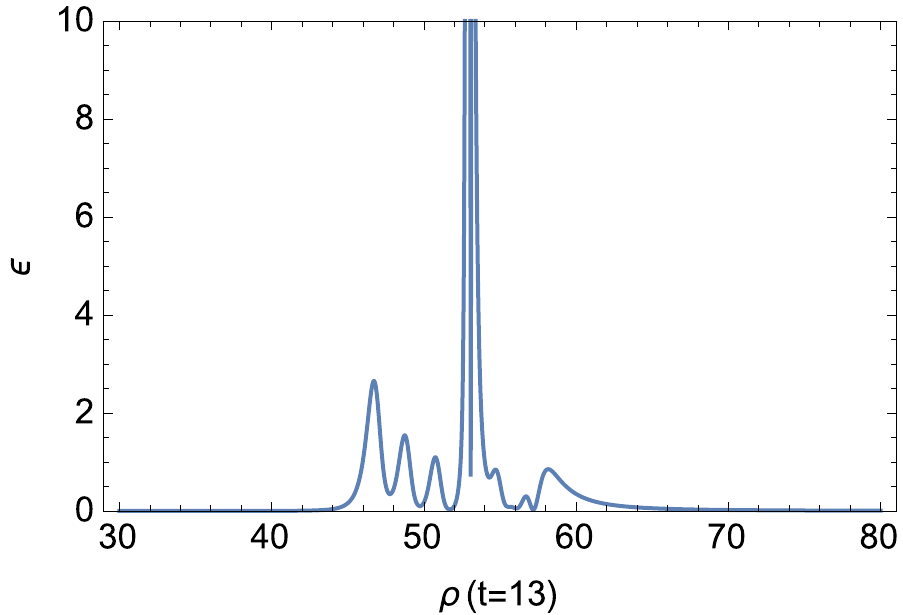}
 \end{minipage} &\ \ \ \ \ \ 
 
  \begin{minipage}[t]{0.3\hsize}
 \centering
\includegraphics[width=5.7cm]{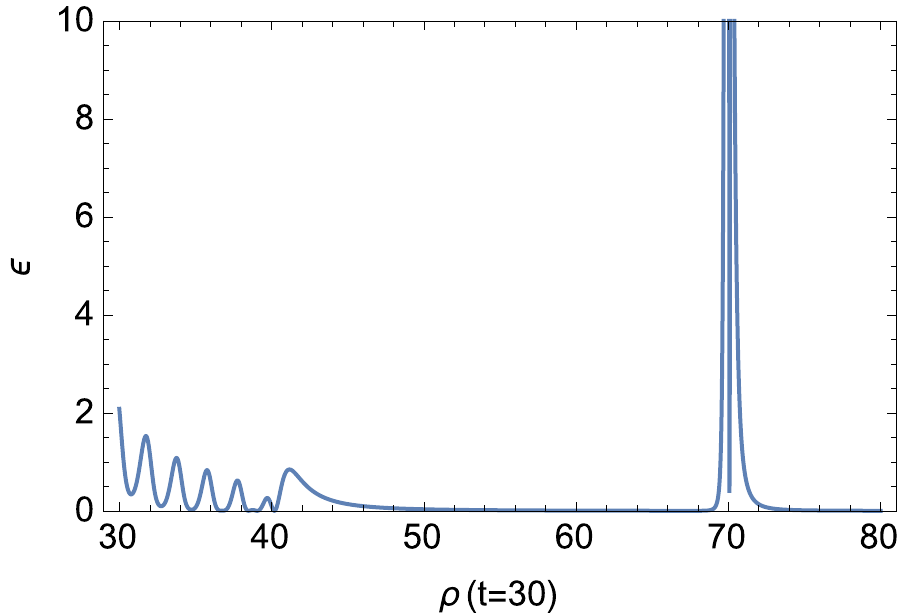}
 \end{minipage}\\ &
 \\

 \begin{minipage}[t]{0.3\hsize}
 \centering
\includegraphics[width=5.7cm]{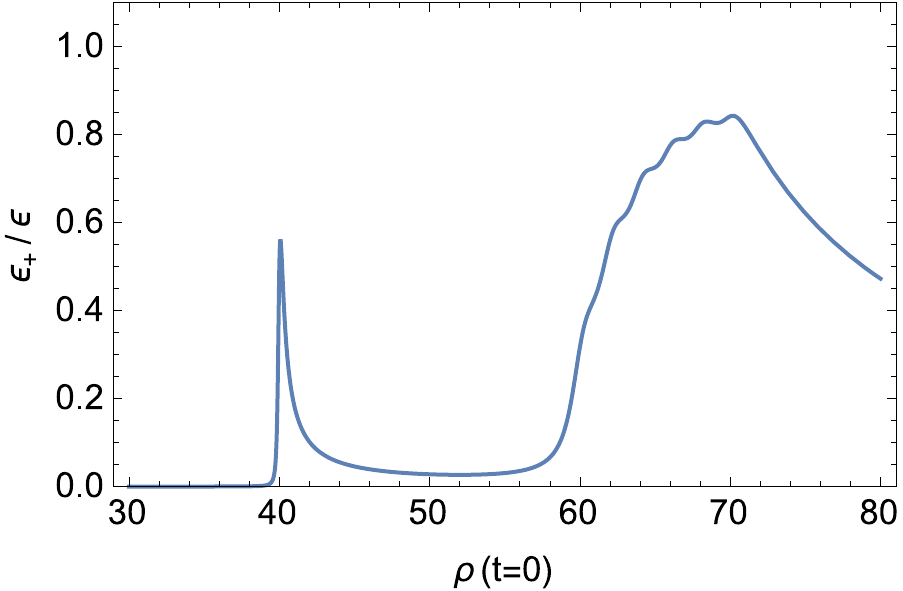}
 \end{minipage}&\ \ \ \ \ \ 
 
 \begin{minipage}[t]{0.3\hsize}
 \centering
\includegraphics[width=5.7cm]{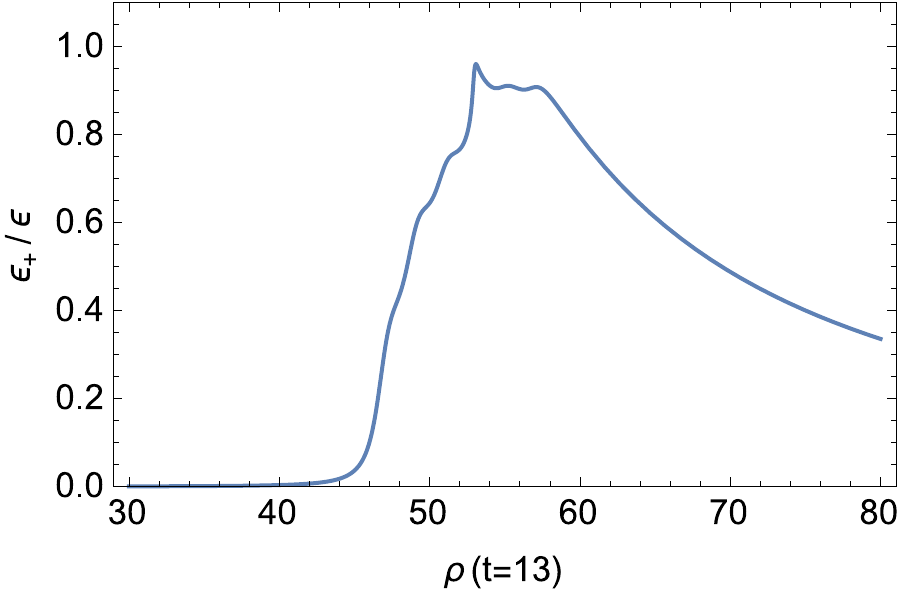}
 \end{minipage} &\ \ \ \ \ \ 
 
  \begin{minipage}[t]{0.3\hsize}
 \centering
\includegraphics[width=5.7cm]{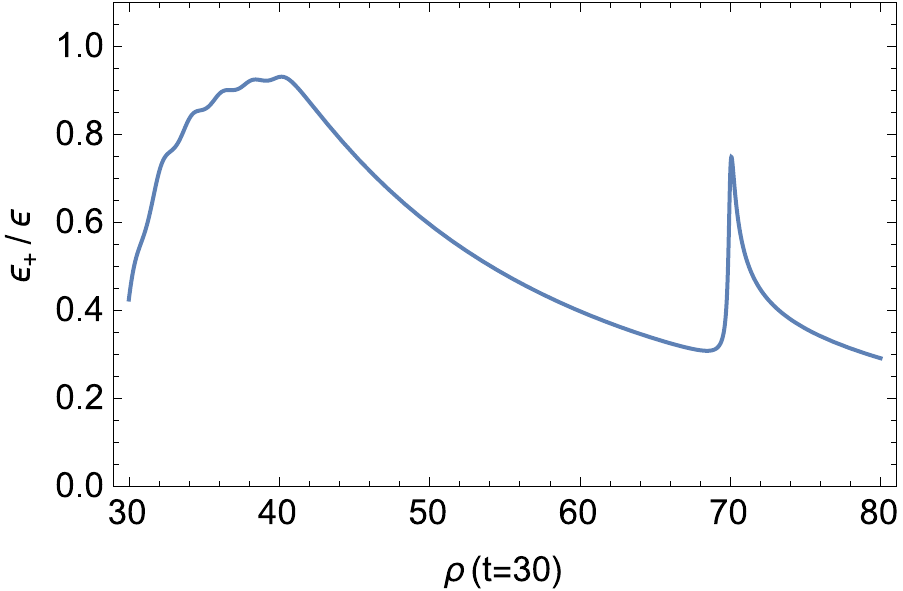}
 \end{minipage}

 \end{tabular}

\caption{The upper graphs show the snapshots at the collision of a single large gravitational pulse with $(t_0,a_0,c_0)=(-40,1/7,2)$ and six small gravitational pulses with $(t_i,a_i,c_i)=(58+2i,1,3)\ (i=1,\cdots,6)$ for $A=1$ and $\delta=0$, where  from left to right, the three graphs correspond to $t=0,13,30$.   The lower graphs show the ratio $\cal E_+/\cal E$ at each time.}
 \label{fig:collision6d}
\end{figure}

\begin{figure}[h]

  \begin{tabular}{ccc}
 \begin{minipage}[t]{0.45\hsize}
 \centering
\includegraphics[width=8cm]{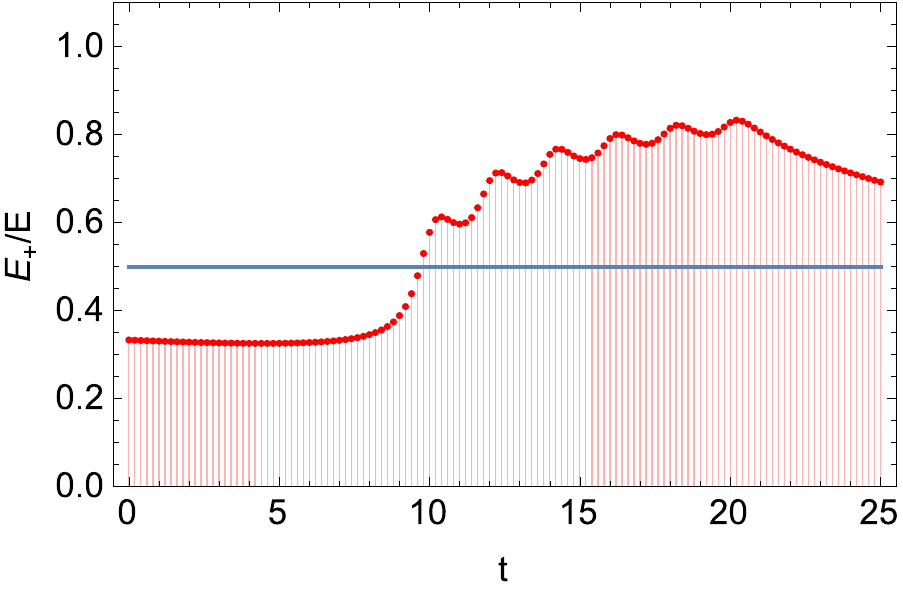}
 \end{minipage} &\ \ \ \ \ \ 
 
 \begin{minipage}[t]{0.45\hsize}
 \centering
\includegraphics[width=8cm]{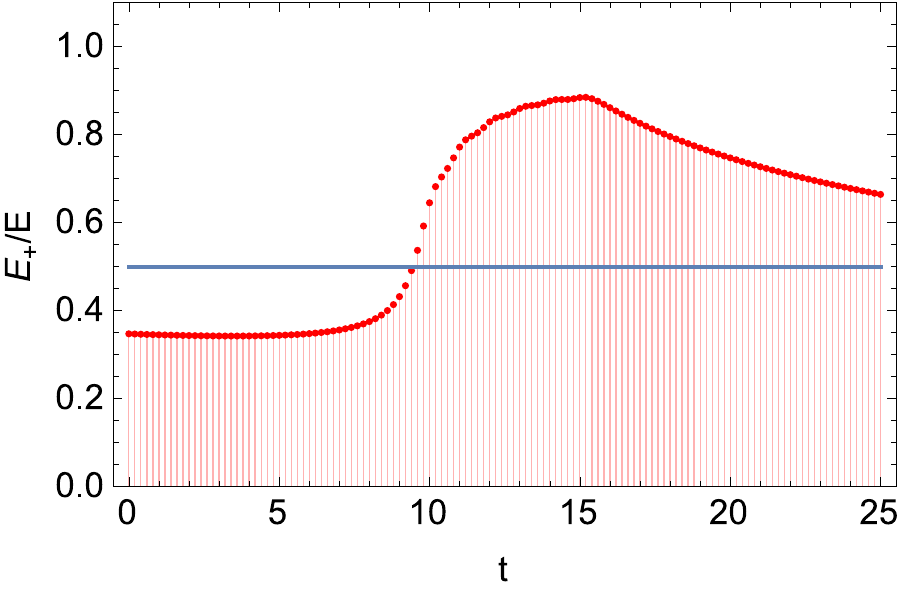}
 \end{minipage} &\ \ \ \ \ \

 \end{tabular}
\caption{ The time dependence of the ratio $E_+(t,1000)/E(t,1000)$ 
 when an outgoing pulse with $(a_0,c_0,t_0)=(1/7,2,-40)$ and six ingoing pulses with $(a_i,c_i)=(1,3)$  collide for $(A=1,0)$, where the left and right graphs display the ratio for $t_i=56+4i$ and   $t_i=58+2i$ $(i=1,\ldots,6)$, respectively.  
}
\label{fig:integ2}
\end{figure}

Figure \ref{fig:collision6c} (Fig.\ref{fig:collision6d}) displays the snapshots of $t=0,13,20$ at the collisions of a large pulse with $(t_0,a_0,c_0)=(-40,1/7,2)$ and six small pulses with $(t_i,a_i,c_i)=(56+4i,1,3)$ [$(t_i,a_i,c_i)=(58+2i,1,3)$ in Fig.\ref{fig:collision6d}] for $A=1$ and  the left (right) graph in Fig.~\ref{fig:integ2} shows the time development of the ratio $E_+(t,1000)/E(t,1000)$ for $0\le t\le 20$. As is shown in these graphs, when their pulses collide, the $\times$ mode is converted to the $+$ mode by the nonlinear effect but it is soon decreased, and after that collision, the $+$ mode becomes dominant at late time.

\medskip
We can summarize as follows.  For $0<A\ll1$, the $+$ mode is dominant at infinity far from the axis of symmetry, whereas during the collision of their pulses, the mode is transformed into the $\times$ mode due to the nonlinear interaction between the pulses.  On the other hand, for $A\simeq 1$, almost only the $\times$ mode exists at infinity, whereas  it is transformed into the $+$ mode at the collision of the pulses.  
Compared with the reflection at the axis of symmetry,  however, for the collision,  the ratio of one mode to the other mode turns out to be smaller.  It may be expected from this result that the more a gravitational wave is concentrated, the greater its nonlinear interaction influences on the mode conversion.


\section{Further generalizations}

In the previous sections, we have studied the generalized Halilsoy pulse solution which is obtained from the generalized  WWB pulse solution. 
Furthermore, we can consider more general pulse solutions that are obtained from the general form of the extended seeds $\tau$,
\begin{eqnarray}
\tau=c\int^\infty_0\left[f(k)e^{-ak}J_0(k\rho)\cos(kt)+g(k)e^{-ak}J_0(k\rho)\sin(kt) \right]dk,
\end{eqnarray}
where we consider that $f(k)$ and $g(k)$ are the arbitrary functions of $k$ such that  this integral converges. 
The generalized Halilsoy pulse solution obtained in the previous sections corresponds to $f(k)=\cos\delta, \ g(k)=\frac{1}{\sqrt{2}}\sin\delta$. In the following, we present three different types of solutions.

\subsubsection{$f(k),\ g(k)$: polynomial functions of $k$  }\label{sec:sol4}
If $f(k)$ and $g(k)$ are polynomial functions of $k$, it is direct to obtain the integral expression of $\tau$ from the derivatives of $\tau_{\rm even}$ and $\tau_{\rm odd}$ with respect to the parameter $a$ as follows:
\begin{eqnarray}
\tau_p&=&cf(-\partial_a)\int^\infty_0 e^{-ak}J_0(k\rho)\cos(kt)dk+cg(-\partial_a)\int^\infty_0e^{-ak}J_0(k\rho)\sin(kt) dk\\
 &=&f(-\partial_a)\tau_{\rm even}+\sqrt{2}g(-\partial_a)\tau_{\rm odd}.
\end{eqnarray}
It can be immediately shown from Eq.(\ref{eq:ratio2}) that the two ratios $\cal{E_+}/\cal{E}$ at future and past null infinities are the same because  $\tau_p$ vanishes at both null infinities, which is easily derived from the fact that $\tau_{\rm even}$ and $\tau_{\rm odd}$ vanish there.

\subsubsection{$f(k)=0,\ g(k)=1/k$ }\label{sec:sol3}
Next, let us consider the case of $f(k)=0$ and $g(k)=1/k$ as a seed solution for the harmonic map, in which case the corresponding pulse solution $ \tau_s$ can be written as
\begin{eqnarray}
\tau_s:=c\int_0^\infty \frac{1}{k}e^{-a k}J_0(k\rho)\sin(kt)dk=c\sin^{-1}\left(\frac{2t}{\sqrt{a^2+(t+\rho)^2}+\sqrt{a^2+(t-\rho)^2}}\right)+\frac{\pi}{2}\alpha,
\end{eqnarray}
where $\alpha$ is a constant and to obtain this integration expression, we have used the formula for the Bessel function in Ref.~\cite{formula}. 
If one replaces $\tau$ in Eqs.(\ref{eq:psi}), (\ref{eq:omega2}) and (\ref{eq:gamma1}) with $\tau_s$, one can get a new pulse solution including the four parameters $(a,c,A,\alpha)$.
At past null infinity $u=-\infty$, the amplitudes behave as  
\begin{eqnarray}
\frac{\rho}{4}A_+^2&\simeq& c^2\left(\frac{1-A^2e^{-2\pi (c-\alpha)}}{1+A^2e^{-2\pi (c-\alpha)}}\right)^2\frac{F(v)}{a^2+4v^2} +{\cal O}((-u)^{-\frac{1}{2}}),\\
\frac{\rho}{4}A_\times^2&\simeq& c^2\left(\frac{2Ae^{-\pi (c-\alpha)}}{1+A^2e^{-2\pi (c-\alpha)}}\right)^2\frac{F(v)}{a^2+4v^2} +{\cal O}((-u)^{-\frac{1}{2}}),
\end{eqnarray}
whereas at future null infinity $v=\infty$, the outgoing amplitude behaves as
\begin{eqnarray}
\frac{\rho}{4}B_+^2&\simeq& c^2\left(\frac{1-A^2e^{2\pi (c+\alpha)}}{1+A^2e^{2\pi (c+\alpha)}}\right)^2\frac{F(-u)}{a^2+4u^2} +{\cal O}(v^{-\frac{1}{2}}),\\
\frac{\rho}{4}B_\times^2&\simeq& c^2\left(\frac{2Ae^{\pi (c+\alpha)}}{1+A^2e^{2\pi (c+\alpha)}}\right)^2\frac{F(-u)}{a^2+4v^2} +{\cal O}(v^{-\frac{1}{2}}).
\end{eqnarray}
It turns out that the ratios of the $\times$ mode to the $+$ mode yield
\begin{eqnarray}
\frac{B_\times^2}{B_+^2}=\frac{4A^2e^{2\pi (c+\alpha)}}{(1-A^2e^{2\pi (c+\alpha)})^2},\quad \frac{A_\times^2}{A_+^2}=\frac{4A^2e^{-2\pi (c-\alpha)}}{(1-A^2e^{-2\pi (c-\alpha)})^2}, 
\end{eqnarray}
which can be immediately obtained from Eqs.~(\ref{eq:ratio}).  
This result implies that the ratio of the $+$ mode to the $\times$ mode for the ingoing pulse from past null infinity is different from that for the outgoing pulse to future null infinity. 
This difference may be considered to be due to the self-interaction at the reflection at the axis. 
It is worth noting that in general, this spacetime is not asymptotically flat because the mathematical analysis in Ref.~\cite{Stachel} shows that $\tau_s$ must vanish at infinity for an asymptotically flat spacetime.

\begin{figure}[h]
  \begin{tabular}{ccc}
 \begin{minipage}[t]{0.3\hsize}
 \centering
\includegraphics[width=5.7cm]{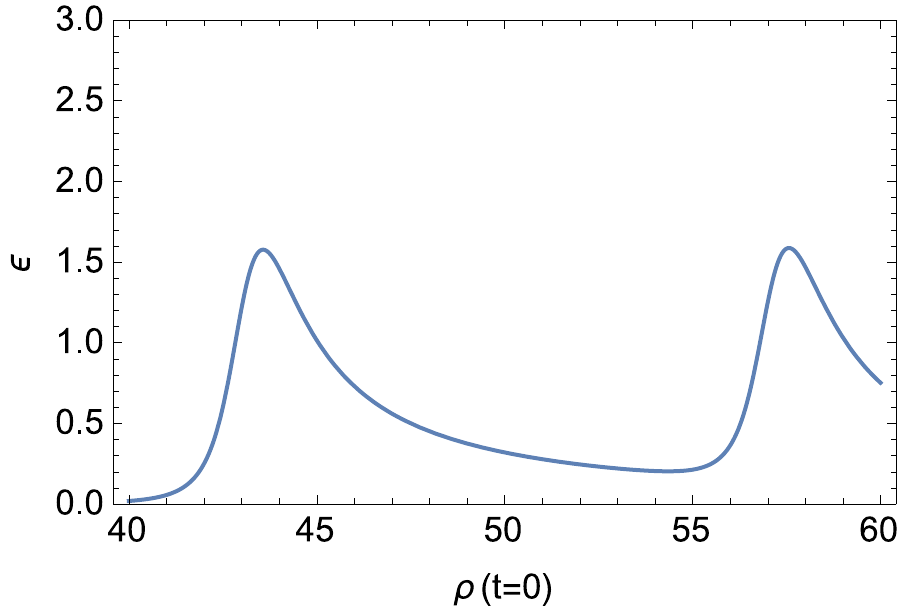}
 \end{minipage} &\ \ \ \ \ \ 
 
 \begin{minipage}[t]{0.3\hsize}
 \centering
\includegraphics[width=5.7cm]{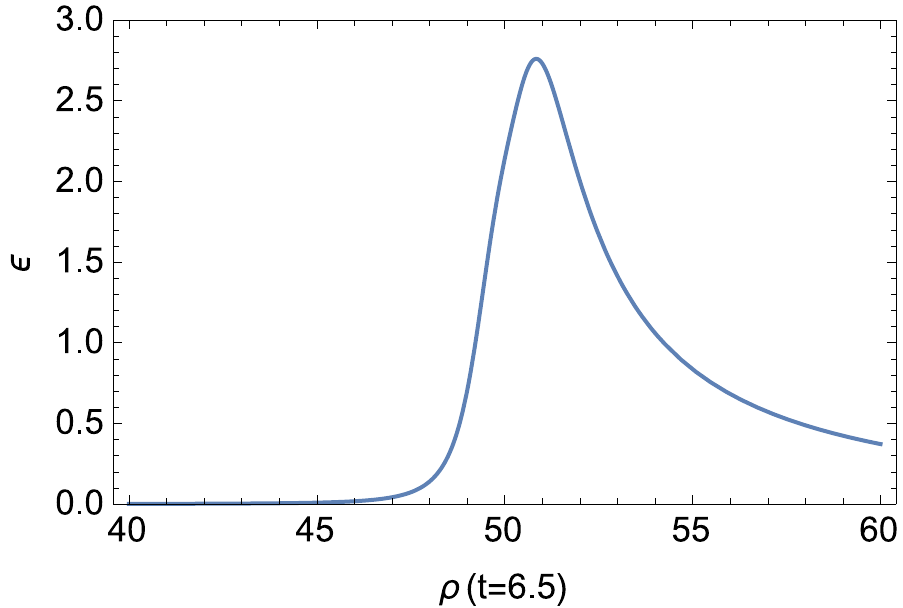}
 \end{minipage} &\ \ \ \ \ \ 
 
  \begin{minipage}[t]{0.3\hsize}
 \centering
\includegraphics[width=5.7cm]{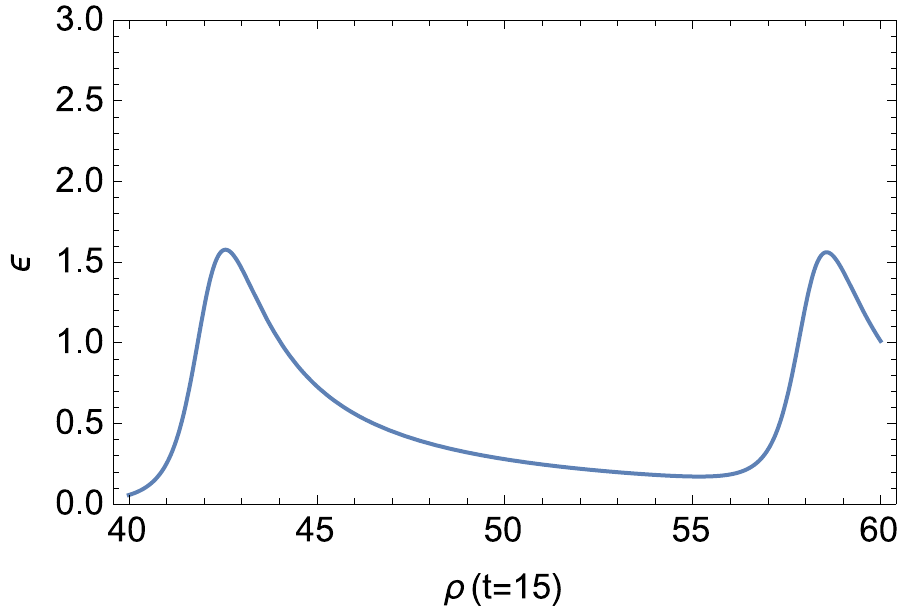}
 \end{minipage}\\ &
 \\

 \begin{minipage}[t]{0.3\hsize}
 \centering
\includegraphics[width=5.7cm]{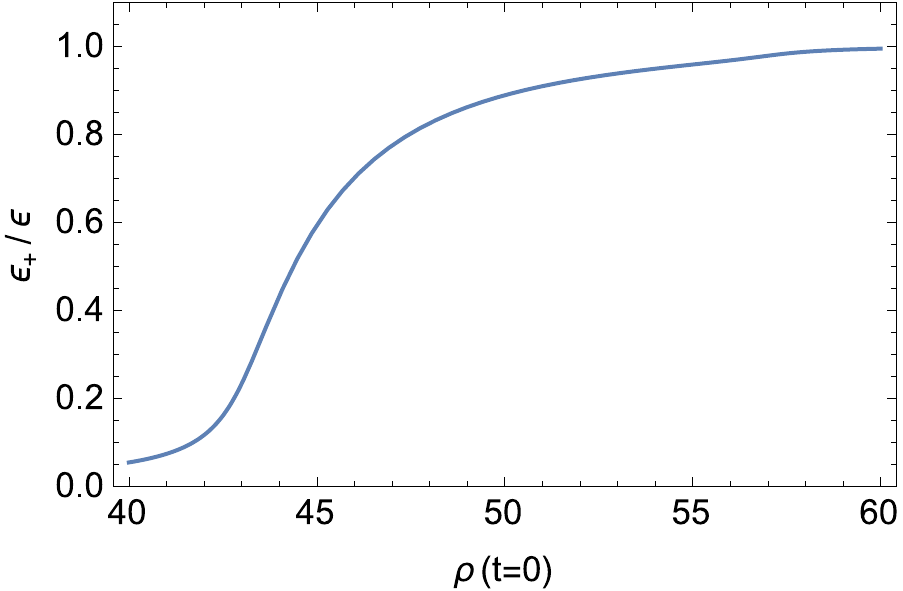}
 \end{minipage}&\ \ \ \ \ \ 
 
 \begin{minipage}[t]{0.3\hsize}
 \centering
\includegraphics[width=5.7cm]{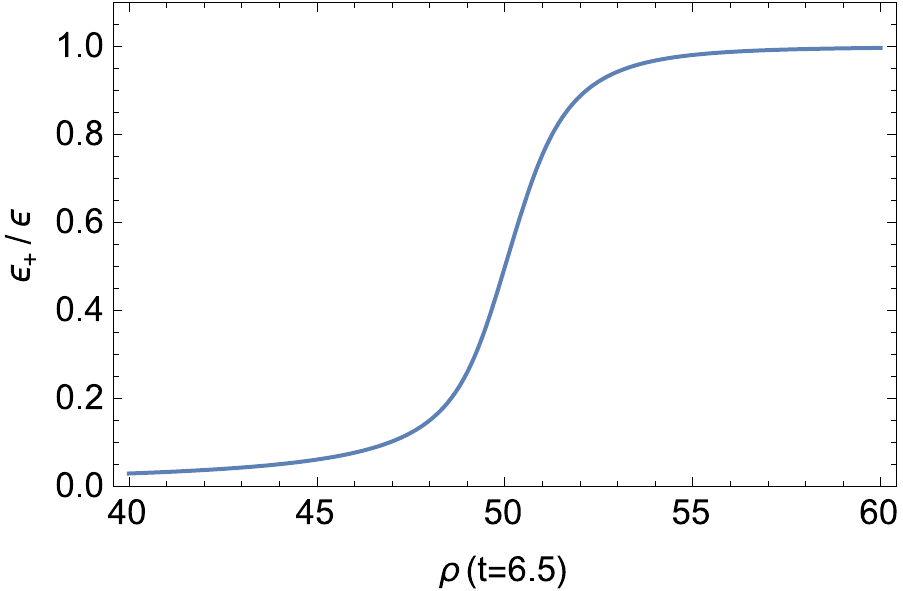}
 \end{minipage} &\ \ \ \ \ \ 
 
  \begin{minipage}[t]{0.3\hsize}
 \centering
\includegraphics[width=5.7cm]{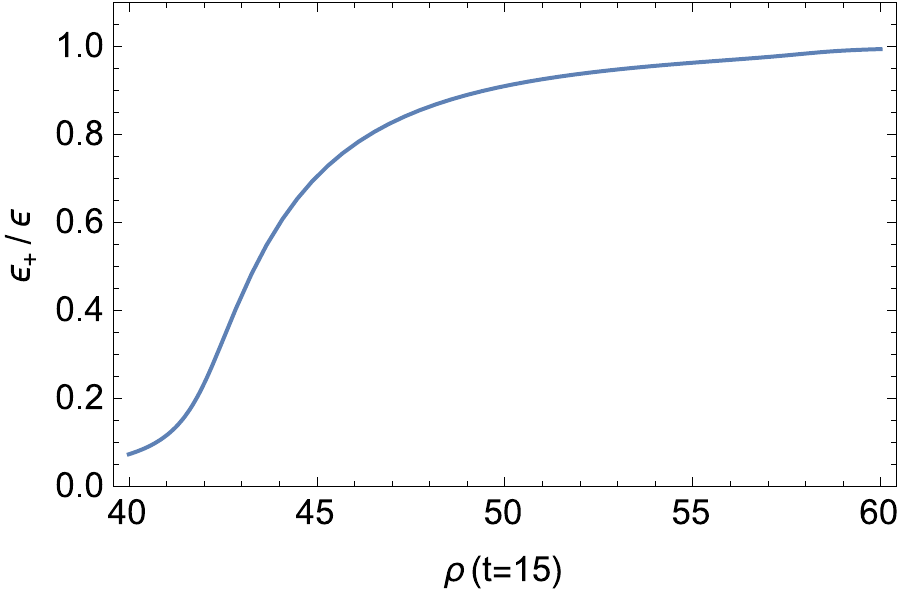}
 \end{minipage}

 \end{tabular}
\caption{The upper graphs show the plots of $\cal E$ at the collision of an outgoing pulse with $(a_1,c_1,\alpha_1)=(1,\pi/2,1)$ and an ingoing pulse with $(a_2,c_2,\alpha_2)=(1,-\pi/2,1)$ for $A=10^{-7}$, where  from left to right, the three graphs correspond to $t=0,6.5,15$.   The lower graphs show the ratio $\cal E_+/\cal E$ at each time.}
 \label{fig:collisionsina}
\end{figure}

\medskip
Furthermore,  as in the previous section, let us consider the collision of two pulses, an outgoing pulse with $(a_1,c_1,\alpha_1,t_1)$ and an ingoing pulse with $(a_2,c_2,\alpha_2,t_2)$ for the same value of $A$.   
 In particular, it is of physical interest to impose the additional condition $c_1=-c_2$ to guarantee asymptotic flatness at null infinity for the solution corresponding to the collision of  the two pulses.  
 The upper figures in Fig.\ref{fig:collisionsina} display the snapshots of $t=0,6.5,10$ at the collision of an outgoing pulse with $(a_1,c_1,\alpha_1,t_1)=(1,\pi/2,1,-43)$ and an ingoing pulse with $(a_2,c_2,\alpha_2,t_2)=(1,-\pi/2,1,57)$ for $A=10^{-7}$. 
As seen in the lower figures, when an outgoing pulse with the $\times$ mode and  an incoming pulse with the $+$ mode collide, they seem to bounce in mutually opposite directions.

\subsubsection{$f(k)=1/\sqrt{k}$, $g(k)=0$ or  $f(k)=0$, $g(k)=1/\sqrt{k}$ }\label{sec:sol2}
Finally, let us consider the case of $f(k)=1/\sqrt{k}$, $g(k)=0$ or  $f(k)=0$, $g(k)=1/\sqrt{k}$. The explicit expression of $\tau$ can be derived by 
using the formula in Ref.~\cite{formula2},
\begin{eqnarray}
\int_0^\infty \frac{1}{\sqrt{k}}e^{-\alpha k}J_0(k\rho)dk=\frac{\sqrt{\pi}}{(\alpha^2+\rho^2)^{\frac{1}{4}}}P^0_{-\frac{1}{2}}\left[\frac{\alpha}{(\alpha^2+\rho^2)^{\frac{1}{2}}}\right],
\end{eqnarray}
and introducing the $(x,y)$ coordinates in~(\ref{eq:trans}),
\begin{eqnarray}
\tau_c&:=c&\int_0^\infty \frac{1}{\sqrt{k}}e^{-a k}e^{ikt}J_0(k\rho)dk\\
&=&c\frac{\sqrt{\pi}}{[(a-it)^2+\rho^2]^{\frac{1}{4}}}P^0_{-\frac{1}{2}}\left[\frac{a-it}{[(a-it)^2+\rho^2]^{\frac{1}{2}}}\right]\\
&=&c\frac{(y+\sqrt{x^2+y^2})^{\frac{1}{2}}+ix(y+\sqrt{x^2+y^2})^{-\frac{1}{2}}}{\sqrt{2a(x^2+y^2)}}P^0_{-\frac{1}{2}}\left[\frac{(1+x^2)y-i(y^2-1)x}{x^2+y^2}\right]\
\end{eqnarray}
where $P^0_{-\frac{1}{2}}(x)$ is  a Legendre function of $x$ and we have put $\alpha=a-it$. From the real and imaginary parts of $\tau_c$, one can obtain the harmonic functions $(\tau_r,\tau_i)$ corresponding to $f(k)=1/\sqrt{k}$, $g(k)=0$ or  $f(k)=0$, $g(k)=1/\sqrt{k}$, namely, as
\begin{eqnarray}
\tau_r:=\Re[{\tau_c]},\quad   \tau_i:=\Im{[\tau_c]}.
\end{eqnarray}
At past null infinity $u\to -\infty$ and future null infinity $v\to\infty$, $v_c$ behaves as, respectively, 
\begin{eqnarray}
\tau_c&\simeq&{\cal O}((-u)^{-\frac{1}{2}}),\\
\tau_c
           &=&{\cal O}(v^{-\frac{1}{2}}).
\end{eqnarray}
Therefore, from~(\ref{eq:ratio}), the obtained solution has
\begin{eqnarray}
\frac{A_\times^2}{A_+^2}=\frac{B_\times^2}{B_+^2}=\frac{4A^2}{(1-A^2)^2}.
\end{eqnarray}


\section{Summary and Discussion}

In this work, we have studied the nonlinear properties of strong gravitational field, based on the analysis of 
the mode conversion of the cylindrically symmetric gravitational waves. 
The solutions that have been treated here are more controllable and of a different type compared with the previous solitonic ones. 
To construct the solutions, we have first extended the WWB pulse solution~\cite{weber:1957,bonnor:1957} within a class of the Einstein-Rosen waves, 
and then, regarding the extended linear pulse solution as a seed, obtained the generalization of the Halilsoy's 
WWB solution with nonaligned polarization in Ref.~\cite{Halilsoy:1988vz} by the harmonic mapping method. 
The general features of the solutions are as follows: the obtained solutions describe the nonlinear gravitational 
pulse waves with the two polarizations ($+$ and $\times$ modes) that come from past null infinity, reflect 
at the axis, and return to future null infinity; among the four parameters that the solutions have, especially 
the parameter $A$ is used to control the extent of the nonlinearity of gravitational waves. 

By varying the parameter $A$, we have observed the time variation of mutual transformation between 
the $+$ mode and the $\times$ mode.
In particular,
through the reflection process of a single pulse wave and collision process of two waves, we have investigated the nonlinearity of the gravitational waves.

To summarize, we can conclude as follows:

\medskip
(i) Reflection: for some parameters (for instance, $A=0.05$), when a pulse with only almost the $+$ mode comes from past null infinity and reflects at the axis, 
it is converted temporarily to the pulse with the $\times$ mode, soon back to the $+$ mode only , and returns to future null infinity, whereas for other ones (for instance, $A=1$), vice versa. 
We may consider that  this occurs due to the very strong nonlinearity since the self-interaction is considerably enhanced when a pulse reflects at the axis of symmetry.

\medskip
(ii) Collision: when gravitational pulses collide, the pulse with only almost the $+$ mode is slightly converted to that of the $\times$ mode but gets back to the $+$ mode only, and vice versa. 
We may consider that the nonlinear interaction has a greater influence on the mode conversion, when gravitational waves are concentrated in a line than when in a planar. In the collision of a single pulse and multipulses, the $+$ mode is more converted, than in the collision of two single pulses. This result implies that the longer gravitational waves with the different modes interact  nonlinearly with each other,  the more one mode is converted to the other mode.

\medskip

In the rest of the summary, we briefly comment on further investigations based on the solution derived here. 
To set physically more interesting problems, it is natural to introduce an appropriate observer (for example, a distant observer) and also some probes interacting with the waves ({\it i.e.}, test particles, test rings, and so on). 
For a distant observer watching the behavior of the waves in the strong interaction region, 
it may be important to know, first of all, whether the conversion phenomenon can be detected or not, 
and, if possible, it may be interesting to determine how the phenomenon take place. 
However, it is easily expected that the observation may be difficult in the case of pure vacuum where only the waves exist, 
because the rate of mode mixing that the waves generally show returns to the initial one quickly after the occurrence of the conversion. 
Once the test objects are introduced we can know the existence and more detailed characteristics of the phenomenon through the motion of those objects. 
In fact, by analyzing the effect of the waves on a test particle, the precedent work~\cite{Bicak:2008gn} shows that the WWB linear polarized wave causes strong ``linear stretching" in the directions transverse to its propagation as well as ``linear dragging" in the radial direction. 
If the conversion occurs at the position where such a particle is placed, it may be expected that the effect of the waves on the particle is qualitatively different from the case of the original WWB wave. 
That is, the ringing way of the particle varies depending on the way of the conversion. 
Such approaches are useful to clarify detailed nonlinear properties of gravitational waves.

\medskip

For a further application, it is noteworthy that the mode conversion of the Einstein vacuum system is 
closely related to that of the Einstein-Maxwell system: 
the mode conversion between two gravitational wave modes can be interpreted as that of 
gravitational waves and electromagnetic waves.  In fact, as is well known~\cite{book exact solution}, the vacuum Einstein equation with cylindrical symmetry is exactly equivalent with the Einstein equation with whole-cylindrical symmetry in the presence of the Maxwell field with only a magnetic potential $A_\phi$.  To see this, let us assume that in the Einstein-Maxwell system, the metric  is written in the diagonal form
\begin{eqnarray}
ds^2=e^{2\tilde\psi}dz^2+\rho^2e^{-2\tilde\psi}d\phi^2+e^{2(\tilde\gamma-\tilde\psi)}(-dt^2+d\rho^2),
\end{eqnarray}
and furthermore, the Maxwell's field strength is denoted by $F=e^{-2\tilde\psi}*(d\tilde \Phi\wedge d\xi)$, where $\tilde \Phi$ is a certain function, and  $\xi=g_{z\mu}dx^\mu$.   Following, for example, Ref.~\cite{Yazadjiev:2005wf}, if one replaces ($\tilde\psi$,$A_\phi$,$\tilde\Phi$,$\tilde\gamma$) with ($2\psi$, $\omega$,$\Phi$,$4\gamma$), respectively, then the Einstein-Maxwell equation coincides with the vacuum Einstein equations~(\ref{eq:Ein1})--(\ref{eq:Ein4}) in the Kompaneets-Jordan-Ehlers form~(\ref{eq:KJE}).
This fact gives us the physically interesting phenomenon that a large portion of the gravitational pulse wave with a single polarization mode can be converted to the electromagnetic pulse wave with a single polarization mode at the axis of symmetry in the same way as a large portion of gravitational pulse with the $+$ mode is converted to that of the $\times$ mode,  as seen in Fig.~\ref{fig:reflection1}.

\medskip
Hence if there exists a charged test particle near the symmetric axis, a distant observer can see easily the way of the conversion 
between the gravitational wave and the electromagnetic wave  through characteristic changes of glittering of the charged particle. 
From the viewpoint of the initial value problem, further, the above fact may lead to a possibility of a strong burst of electromagnetic waves: 
if the concentration of gravitational waves with the $+$ mode exists near the axis at the initial time, the gravitational waves decay immediately, and at the same time, the strong electromagnetic waves are generated with only the little electromagnetic wave existing. 
Further analysis on this point will also be interesting.

\medskip
Finally, we briefly comment on the alternative cylindrical gravitational energy mentioned in the introduction and compare it with the C-energy. 
From a standpoint of symmetry reduction, the other definitions of the energy were given in Ref.~\cite{Ashtekar:1997}  for (the Einstein-Rosen type of) a diagonal metric and  in Ref.~\cite{Sjodin:2000zd}  for an off-diagonal metric, respectively. 
According to them, the energy density is defined by
\begin{eqnarray}
{\cal \tilde E}= \frac{\rho}{8}{e^{-\gamma}}\gamma_{,\rho}=\frac{\rho}{8}{e^{-\gamma}}(A_+^2+B_+^2+A_\times^2+B_\times^2),
\end{eqnarray}
which leads us to the following natural definitions for the energy densities corresponding to the $+$ and $\times$ modes,
\begin{eqnarray}
{\cal \tilde E}_+&=& \frac{\rho}{8}e^{-\gamma}(A_+^2+B_+^2),\\
{\cal \tilde E}_\times &=& \frac{\rho}{8}e^{-\gamma}(A_\times^2+B_\times^2),
\end{eqnarray}
where  one notes that the factor $e^{-\gamma}$ comes from the volume element of the region symmetry reduced along the $z$ direction. 
The total energy  contained within the radius $\rho_0$ (: constant) at a certain time $t$, the total energies assigned to the $+$ and $\times$ modes are defined as, respectively, 
\begin{eqnarray}
\tilde E(t,\rho_0)&=&\int_0^{\rho_0}{\cal \tilde E}(t,\rho)d\rho,\\ 
\tilde E_+(t,\rho_0)&=&\int_0^{\rho_0}{\cal \tilde E}_+(t,\rho)d\rho,\\ 
\tilde E_\times(t,\rho_0)&=&\int_0^{\rho_0}{\cal \tilde E}_\times(t,\rho)d\rho.
\end{eqnarray}
The behaviors of this ratio ${\cal \tilde E}_+/{\cal \tilde E}$ are exactly the same as that of the ratio $\cal E_+/\cal E$ defined in terms of the C-energy densities since this extra factor $e^{-\gamma}$ is reduced. 
Moreover, in general, the ratio of total energy $\tilde E_+(t,\rho_0)/\tilde E(t,\rho)$ does not coincide with the ratio $E_+(t,\rho_0)/ E(t,\rho)$ but  we have numerically checked that there there is no qualitative difference from the results in this paper.

\acknowledgments
This work was supported by the Grant-in-Aid for Young Scientists (B) (Grant. Number~26800120) and Grant-in-Aid for Scientific Research (C) (Grant Number ~17K05452) from theJapan Society for the Promotion of Science (S.T.).

\end{document}